\documentclass{article}
\usepackage{jheppub} 
\usepackage[utf8x]{inputenc}	
\usepackage[english]{babel}

\usepackage[ddmmyyyy,hhmmss]{datetime}

\usepackage{multirow}
\catcode`@=11
\allowdisplaybreaks
\usepackage{amsxtra}
\usepackage{mathtools}
\usepackage{float}
\usepackage{xcolor}
\usepackage{tikz}
\usepackage{stackengine}
\usepackage{placeins}
\usepackage{booktabs}
\usepackage{macros}
\usepackage{hyperref}
\usepackage{subcaption}
\usepackage{amsmath}
\usepackage{graphicx}
\usepackage{rotating}
\usepackage{epigraph}

\usetikzlibrary{hobby,intersections,positioning} 
\usetikzlibrary{shapes.geometric}
\newcommand\sbullet[1][.75]{\mathbin{\vcenter{\hbox{\scalebox{#1}{$\bullet$}}}}}

\newcommand{\Figref}[1]{Figure~\ref{#1}}

\newcommand{\pe}{\mathrm{PE}}

\usepackage{placeins}
\usetikzlibrary{calc}
\usetikzlibrary{arrows.meta, bending, patterns}

\tikzset{flavour/.style={draw=none,minimum size=0.3mm,fill=white, regular polygon,regular polygon sides=4,draw}}
\tikzset{flavourr/.style={draw=none,minimum size=0.3mm,fill=red, regular polygon,regular polygon sides=4,draw}}
\tikzset{flavourb/.style={draw=none,minimum size=0.3mm,fill=blue, regular polygon,regular polygon sides=4,draw}}
\tikzset{gaugeBig/.style={inner sep=1mm,draw=none,fill=white,minimum size=2mm,circle, draw}}
\tikzset{bd/.style={circle, draw=black, inner sep=0pt, fill=black, minimum size=2mm}}
\tikzset{wd/.style={circle, draw=black, inner sep=0pt, fill=white, minimum size=2mm}}
\tikzset{Dynkin/.style={circle, draw=black, inner sep=0pt, fill=white, minimum size=2mm}}
\tikzstyle{ligne}=[draw, very thick] 
\tikzstyle{gridline}=[draw, gray] 
\tikzset{gauge/.style={circle, draw,inner sep=2.5pt}}
\tikzset{gaugeo/.style={circle, draw,inner sep=2.5pt,fill=orange}}
\tikzset{gauger/.style={circle, draw,inner sep=2.5pt,fill=red}}
\tikzset{gaugeb/.style={circle, draw,inner sep=2.5pt,fill=blue}}
\tikzset{gaugeg/.style={circle, draw,inner sep=2.5pt,fill=green}}
\tikzset{gaugegoodgreen/.style={circle, draw,inner sep=2.5pt,fill=goodgreen}}
\tikzset{gaugem/.style={circle, draw,inner sep=2.5pt,fill=magenta}}
\tikzset{hasse/.style={circle, fill,inner sep=2pt}}
\tikzset{d2/.style={circle, fill,inner sep=1.3pt}}
\tikzset{shrinky/.style={circle, fill,inner sep=1pt}}
\tikzset{sized/.style={circle, draw, inner sep=1.5pt}}
\tikzset{seven/.style={circle, draw,inner sep=3pt}}

\tikzset{gaugebl/.style={circle,draw=black,fill=black,inner sep=1.5pt}}
\tikzset{gaugeblnormal/.style={circle,draw=black,fill=black,inner sep=2.5pt}}
\tikzset{hasse/.style={circle, fill,inner sep=2pt}}
\tikzstyle{dashed_brane}=[thick, dashed]
\tikzstyle{dotted_brane}=[thick, dotted]
\tikzstyle{O3plus}=[thick, color=green]
\tikzstyle{O3minustilde}=[thick, color=blue]
\tikzstyle{O3plustilde}=[thick, color=red]
\tikzset{D5/.style={cross out, draw=black, minimum size=7, inner sep=0pt, outer sep=0pt}, cross/.default={1pt}}
\tikzset{flavor/.style={regular polygon,regular polygon sides=4,inner sep=2.5pt, label = {}, draw}}
\tikzset{redflavor/.style={regular polygon,regular polygon sides=4,inner sep=2.5pt, color=red, label = {}, draw}}
\tikzset{redgauge/.style={inner sep=1mm,color=red,draw=none,minimum size=2mm,circle, draw}}
\tikzset{blueflavor/.style={regular polygon,regular polygon sides=4,inner sep=2.5pt, color=blue, label = {}, draw}}
\tikzset{bluegauge/.style={inner sep=1mm,color=blue,draw=none,minimum size=2mm,circle, draw}}
\allowdisplaybreaks

\usepackage{todonotes}

\title{$\mathrm{O}$/$\mathrm{SO}$ Gauge Groups, $BC$ Quivers and $O3$ Planes}

\author{Sam Bennett}
\author{, Amihay Hanany}

\affiliation{Abdus Salam Centre for Theoretical Physics, Imperial College London,\\ Prince Consort Road
London, SW7 2AZ, UK}
\emailAdd{samuel.bennett18@imperial.ac.uk}
\emailAdd{a.hanany@imperial.ac.uk}

\preprint{Imperial/TP/22/AH/01}
\abstract{
D3-/D5-/NS5-brane systems with $O3$ orientifold planes realise 3d $\mathcal{N}=4$ gauge theories with orthogonal and symplectic gauge groups on the D3-brane worldvolume. Such setups have long contained an ambiguity regarding the global form of $D$-type gauge groups. This note offers a partial prescription for reading $\orm(2k)$ and $\sorm(2k)$ gauge nodes in orthosymplectic quivers using the presence of $\frac{1}{2}$D5-branes on the orientifolds bordering $\frac{1}{2}$NS5-brane intervals spanned by $O3^{-}$ planes. A set of identities are proposed relating the Coulomb branches of generic quivers under a Higgsing that relates $\sorm(2k+1)$ and $\orm(2k)$ gauge groups. A further prescription is conjectured regarding the action of $\frac{1}{2}$D5-branes on maximal $DC$-chains.}
\begin{document}
\maketitle
\flushbottom
\section{Introduction}
An interesting feature of the recent history of unitary quiver gauge theories with eight supercharges is the extent to which a surprisingly general set of algorithms captures the essential data of the moduli space \cite{Cabrera:2018ann, Gledhill:2021cbe, Bourget:2023dkj, Cabrera:2016vvv, Cabrera:2017njm, Hwang:2021ulb, Comi:2022aqo, Giacomelli:2023zkk, Giacomelli:2024laq, Grimminger:2024mks, Bennett:2024loi, Bennett:2024llh}. This builds upon the simple fact that the moduli space of a 3d $\mathcal{N}=4$ gauge theory is a symplectic singularity in the sense of Beauville \cite{beauville}, and as such admits a symplectic stratification into so-called `minimal degenerations' whose partial order can be captured using a Hasse\footnote{Symplectic singularity structure emerges from the supersymmetry --- the moduli space of a 3d $\mathcal{N}=2$ theory is often a Gorenstein singularity, whose stratification is far less studied.} diagram \cite{Bourget:2019aer, Bourget:2022ehw, Bourget_2022dim6, Grimminger:2020dmg}. In 3d $\mathcal{N}=4$ theories, scalar fields in the hypermultiplets and vectormultiplets partition the total moduli space into two hyper-K\"ahler spaces known as the Higgs and Coulomb branches, parametrised by the VEVs of each set of scalars respectively. Although the Higgs branch is classically exact on account of the nonrenormalisation theorem, the Coulomb branch is corrected in the quantum theory through the contributions of monopole operators \cite{Borokhov:2002ib, Borokhov:2002cg}, parametrised by the Weyl orbits of the weight lattice of the Langlands dual of the gauge group. Despite these complexities, the rigid symplectic holomorphic structure imposed by the supersymmetry encodes the symplectic stratification of the Coulomb and Higgs branches into the quivers themselves --- the quiver subtraction programme has over the past few years developed several methods for the calculation of Hasse diagrams from quiver combinatorics.

Although Higgs branch quiver subtraction \cite{Bennett:2024loi} can be derived purely from the classical Higgs mechanism (or equivalently by interpreting the Higgs branch as the hyper-K\"ahler quotient of the moduli space of representations of the quiver \footnote{We thank Gwyn Bellamy and Travis Schedler for communicating their results.}), quantum corrections on the Coulomb branch require further information, such as from a brane system, in order to capture the full picture; for unitary simply-laced quivers, such is provided using configurations of D3-branes, D5-branes and NS5-branes in the tradition of Hanany-Witten \cite{Hanany:1996ie}. The inclusion of orientifold planes in D3-/D5-/NS5-brane systems \cite{Hanany:1999sj,Hanany_2001}, which results in orthosymplectic and non-simply laced 3d $\mathcal{N}=4$ quivers arising as low-energy worldvolume theories on the D3-branes \cite{Feng_2000, Hanany_2001}, complicates the analysis significantly. Roughly, the orientifold planes move the natural reference point for the moduli spaces of such theories from the nilpotent cone of $\mathfrak{sl}_{n}$ to those of the classical algebras of types $B/C/D$, introducing new features such as non-normality, non-special orbits and the canonical quotient of Lusztig \cite{2023arXiv231000521J, Generic_singularities, SOMMERS2001790}.

One of the most fundamental obstacles to realising an algorithm for Coulomb branch quiver subtraction on framed orthosymplectic quivers is the problem of determining the global form of the gauge group. The Chan-Paton prescription \cite{MARCUS198297} specifies the gauge group of the resulting theory on a stack of D-branes in the presence of an orientifold, favouring $\orm(n)$ rather than $\sorm(n)$ gauge groups. In quiver gauge theory, a lack of tools for dealing with discrete groups leaves an ambiguity between $\sorm(2N)$ and $\orm(2N)$ gauge theories in NS5-brane intervals spanned by $O3^{-}$ planes. Recent work \cite{Cabrera:2017ucb} on framed orthosymplectic quivers using the framework of nilpotent cones has emphasised the importance of $\sorm(2N)$ versus $\orm(2N)$ --- gauging an extra $\mathbb{Z}_{2}$ in general changes the Coulomb branch (its effect on the Higgs branch was considered in \cite{Cabrera:2018ldc, Bourget:2017tmt, Hanany:2016gbz}).

This note conjectures a rule for reading individual $\orm(2n)$ and $\sorm(2n)$ gauge nodes from D3/D5/NS5-brane systems with $O3$ orientifold planes using the presence of $\frac{1}{2}$D5-branes in neighbouring $\frac{1}{2}$NS5-brane intervals. This rule is supported by various Hilbert series calculations matching moduli spaces across 3d mirror symmetry, which is realised as S-duality \cite{Hanany:1996ie} in the brane systems under consideration. 

Somewhat in parallel is the introduction of a set of identities between the Coulomb branches of a family of framed orthosymplectic gauge theories. Under certain conditions, these swap gauge nodes of $B$-type with those of type $\orm(2n)$ (along with a compensating change in flavour nodes). Using the aforementioned $\orm(2n)$ vs $\sorm(2n)$ rule, these identities are conjectured to have the interpretation of a Higgsing procedure that leaves the Coulomb branch invariant --- in a brane system, this can be identified as a Kraft-Procesi transition on the $\frac{1}{2}$D3-branes created when a D5-brane splits on an $\widetilde{O3^{-}}$ plane \cite{Feng_2000}. From the point of view of the Higgs branch Hasse diagram, this transition is conjectured to occur between two leaves that form part of the same special piece. \footnote{In fact, it implies a stronger statement. Namely, that the two leaves map to the same leaf under symplectic duality.}

This note also conjectures an identity relating the Coulomb branches of the `$BC$-' and `$CB$-chain' quivers, defined in Section \ref{sec:BC_Chains}. Like the other conjectures in this note, this identity is supported via an explicit monopole formula calculation at the level of the unrefined Hilbert series. This Coulomb branch map can be performed in conjunction with those of Section \ref{sec:Bn_O2n_Identity}.

Lastly, this note introduces a further conjecture regarding magnetic lattice gaugings and their emergence from brane systems. Using a conjecture regarding the behaviour of $\mathbb{Z}_{2}$ flavour symmetry gaugings in $\sprm(k)$ SQCD under 3d mirror symmetry, a rule is introduced for reading diagonal $\mathbb{Z}_{2}$ quotients on magnetic lattices from D3-/D5-/NS5-brane systems with $O3$ orientifold planes, which is checked explicitly.

\section{Brane and Quiver Conventions}
\label{sec:conventions}
The brane systems in this note consist of D3-branes suspended between D5- and NS5-branes in the presence of $O3$ orientifold planes. Their spacetime occupancy, given in \Figref{Fig:brane_setup}, leaves the threebranes free to extend through the $x^{6}$ dimension while the D5- and NS5-branes are infinite in the $x^{3,4,5}$ and $x^{7,8,9}$ directions respectively.
\begin{figure}[h]
    \centering
\begin{tabular}{ccccccccccc}
  \hline
  \\[-1em]
   & $x^{0}$ & $x^{1}$ & $x^{2}$ & $x^{3}$ & $x^{4}$ & $x^{5}$ & $x^{6}$ & $x^{7}$ & $x^{8}$ & $x^{9}$\\
  \hline
  NS5 & $\sbullet$ & $\sbullet$ & $\sbullet$ & $\sbullet$ & $\sbullet$ & $\sbullet$ & & & &\\ 
  \hline
  D5 & $\sbullet$ & $\sbullet$ & $\sbullet$ & & & & & $\sbullet$ & $\sbullet$ & $\sbullet$\\ 
  \hline
  D3/$O$3 & $\sbullet$ & $\sbullet$ & $\sbullet$ &  & & & $\sbullet$ &  & & \\ 
  \hline
\end{tabular}
\caption{The Type IIB configurations in this note consist of D3-branes suspended between D5-branes and/or NS5-branes, together with orientifold planes at the origin.}
\label{Fig:brane_setup}
\end{figure}
In brane diagrams, the various $O3$ planes will be notated as in Table \ref{orientifold_table} --- $O3^{\pm}$ are exchanged across $\frac{1}{2}$NS5-branes while $\frac{1}{2}$D5-branes interchange tildered and untildered orientifolds. $S$-duality interchanges $O3^{+}$ and $\widetilde{O3^{-}}$, leaving $O3^{-}$ and $\widetilde{O3^{+}}$ untouched. Further information regarding the brane configurations used in this paper can be found in \cite{Feng_2000, Cabrera:2017ucb, Cabrera:2016vvv, Cabrera:2017njm}.
\begin{table}[H]
\ra{1.5}
    \centering
    \begin{tabular}{cccc}
    \toprule
          Orientifold Plane & Brane Diagram & Electric Gauge Algebra & Magnetic Gauge Algebra\\ \midrule
          $O3^{+}$ & \begin{tikzpicture}
              \draw[dashed_brane] (-1,0) -- (1,0);
          \end{tikzpicture} & $\mathfrak{c}$ & $\mathfrak{b}$ \\ \midrule
          $\widetilde{O3^{+}}$ & \begin{tikzpicture}
              \draw[dotted_brane] (-1,0) -- (1,0);
          \end{tikzpicture} & $\mathfrak{c}$ & $\mathfrak{c}$\\ \midrule
          $\widetilde{O3^{-}}$ & \begin{tikzpicture}
              \draw[-] (-1,0) -- (1,0);
          \end{tikzpicture} & $\mathfrak{b}$ & $\mathfrak{c}$\\ \midrule
          $O3^{-}$ & & $\mathfrak{d}$ & $\mathfrak{d}$
          \\\bottomrule
    \end{tabular}
    \caption{The identification of gauge algebras from $O3$-planes and brane diagram conventions. Switching from the electric to magnetic gauge algebra involves S-duality.}
    \label{orientifold_table}
\end{table}
\subsection{Quiver Derivation Rules}
\label{Section:Derivation_Rules}
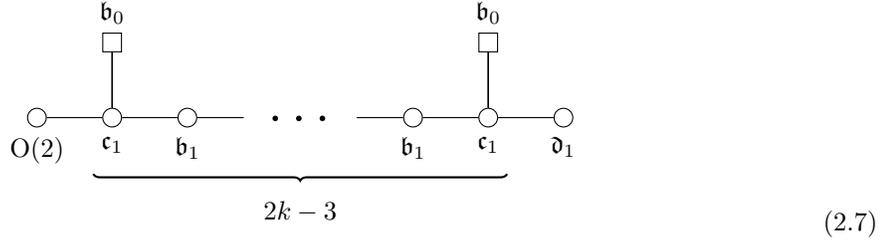
\begin{figure}[h!]
    \centering
    \begin{subfigure}{0.49\textwidth}
    \centering
    \begin{tikzpicture}
        \draw[-] (0,0) -- (0,2);
        \draw[-] (2,0)--(2,2);
        \draw[-] (4,0)--(4,2);
        \draw[-] (-2,1) -- (0,1);
        \draw[dotted_brane] (0,1)--(1,1); 
        \draw[dashed_brane] (1,1)--(2,1); 
        \node (g1) at (1,1) [D5] {};
        \node (g2) at (2.4,1) [D5] {};
        \node (g3) at (2.8,1) [D5] {};
        \node (g4) at (3.2,1) [D5] {};
        \node (g5) at (3.6,1) [D5] {};
        \draw[-] (2.4,1)--(2.8,1);
        \draw[-] (3.2,1)--(3.6,1);
        \draw[-] (0,1.8)--(2,1.8);
        \draw[-] (0,0.2)--(2,0.2);
        \draw[-] (2,0.6)--(4,0.6);
        \draw[-] (2,1.4)--(4,1.4);
    \end{tikzpicture}
    \caption{}
    \label{fig:semi_inf_O3minustilde_plane_1}
    \end{subfigure}
    \begin{subfigure}{0.49\textwidth}
    \centering
    \begin{tikzpicture}
        \node (1) [gauge, label=below:{$\mathfrak{c}_{1}$}] at (0,0) {};
        \node (2) [gauge, label=below:{$\mathfrak{d}_{2}$}] at (1,0) {};
        \node (3) [flavor, label=above:{$\mathfrak{c}_2$}] at (1,1) {};
        \node (4) [flavor, label=above:{$\mathfrak{b}_{0}$}] at (0,1) {};
        \draw (4)--(1)--(2)--(3);
        \end{tikzpicture}
    \caption{}
    \label{fig:semi_inf_O3minustilde_plane_2}
    \end{subfigure}
    \begin{subfigure}{0.49\textwidth}
    \centering
    \begin{tikzpicture}
        \draw[-] (0,0) -- (0,2);
        \draw[-] (2,0)--(2,2);
        \draw[-] (4,0)--(4,2);
        \draw[dashed_brane] (0,1)--(0.66,1); 
        \draw[dotted_brane] (0.66,1)--(1.32,1); 
        \draw[dashed_brane] (1,1)--(2,1); 
        \node (g1) at (0.66,1) [D5] {};
        \node (g11) at (1.32,1) [D5] {};
        \node (g2) at (2.4,1) [D5] {};
        \node (g3) at (2.8,1) [D5] {};
        \node (g4) at (3.2,1) [D5] {};
        \node (g5) at (3.6,1) [D5] {};
        \draw[-] (2.4,1)--(2.8,1);
        \draw[-] (3.2,1)--(3.6,1);
        \draw[-] (0,1.8)--(2,1.8);
        \draw[-] (0,0.2)--(2,0.2);
        \draw[-] (2,0.6)--(4,0.6);
        \draw[-] (2,1.4)--(4,1.4);
    \end{tikzpicture}
    \caption{}
    \label{fig:semi_inf_O3minustilde_plane_3}
    \end{subfigure}
    \begin{subfigure}{0.49\textwidth}
    \centering
    \begin{tikzpicture}
        \node (1) [gauge, label=below:{$\mathfrak{c}_{1}$}] at (0,0) {};
        \node (2) [gauge, label=below:{$\mathfrak{d}_{2}$}] at (1,0) {};
        \node (3) [flavor, label=above:{$\mathfrak{c}_2$}] at (1,1) {};
        \node (4) [flavor, label=above:{$\mathfrak{d}_{1}$}] at (0,1) {};
        \draw (4)--(1)--(2)--(3);
        \end{tikzpicture}
    \caption{}
    \label{fig:semi_inf_O3minustilde_plane_4}
    \end{subfigure}
    \caption{The brane system of \Figref{fig:semi_inf_O3minustilde_plane_1} does not give the anomalous theory in \Figref{fig:semi_inf_O3minustilde_plane_2}. After bringing in a $\frac{1}{2}$D5-brane from infinity and performing an HW-transition, obtaining the brane system in \Figref{fig:semi_inf_O3minustilde_plane}, the non-anomalous quiver \Figref{fig:semi_inf_O3minustilde_plane_4} is read.}
    \label{fig:semi_inf_O3minustilde_plane}
\end{figure}

Given a D3-/D5-/NS5-brane system described in the previous section, the following rules identify the orthosymplectic quiver corresponding to the low energy worldvolume theory on the D3-branes. Table \ref{orientifold_table} is of course well-known and is included only for completeness. (Similarly, rules regarding flavours are widely used in \cite{Cabrera:2017njm, Feng_2000}.)
\begin{itemize}
    \item Given an interval between two $\frac{1}{2}$NS5-branes containing both an $O3^{+}/\widetilde{O3^{+}}/\widetilde{O3^{-}}$ plane and a collection of D3-branes, the gauge algebra of the corresponding quiver gauge node is of the form given in Table \ref{orientifold_table}.
    \item In general, $\frac{1}{2}$D5-branes on the $O3$ orientifold planes contribute flavours in the corresponding quiver. The diagrams below give derivation rules for gauge nodes and flavours for D3-branes on various $O3$ orientifold planes. The dictionary given by \eqref{rule_C_gauge_B0_flavour}, \eqref{rule_C_gauge_B_flavour}, \eqref{rule_C_gauge_D_flavour} and \eqref{rule_B_gauge_C_flavour} is well-known from \cite{Feng_2000, Cabrera:2017njm}.
    \begin{equation}
        \centering
        \begin{tikzpicture}
            \node (a) at (0,0) {$\begin{tikzpicture}
                \draw[-] (2,1)--(2.5,1);
                \draw[-] (-2,0) -- (-2,2);
                \draw[-] (2,0) -- (2,2);
                \draw[-] (-2,1.75) -- (2,1.75);
                \draw[-] (-2,0.25) -- (2,0.25);
                \node (g1) at (0,1) [D5] {};
                \draw [dashed_brane] (-2,1) -- (0,1);
                \draw [dotted_brane] (0,1) -- (2,1);
                \node at (0,2) {$\ell$};
            \end{tikzpicture}$};
            \node (b) at (4,0) {$\begin{tikzpicture}
                \node at (0,0) [scale=1]{$\longleftrightarrow$};
            \end{tikzpicture}$};
            \node (c) at (6,0) {$\begin{tikzpicture}
                \node (1) [gauge, label=below:{$\mathfrak{c}_{\ell}$}] at (0,0) {};
                \node (2) [flavor, label=above:{$\mathfrak{b}_{0}$}] at (0,1) {};
                \draw[-] (1)--(2);
            \end{tikzpicture}$};
        \end{tikzpicture}
        \label{rule_C_gauge_B0_flavour}
    \end{equation}
    \begin{equation}
        \centering
        \begin{tikzpicture}
            \node (a) at (0,0) {$\begin{tikzpicture}
                \draw[-] (-2,1)--(-2.5,1);
                \draw[-] (-2,0) -- (-2,2);
                \draw[-] (2,0) -- (2,2);
                \draw[-] (-2,1.75) -- (2,1.75);
                \draw[-] (-2,0.25) -- (2,0.25);
                \node (g1) at (-1.5,1) [D5] {};
                \node (g2) at (-1,1) [D5] {};
                \node (g3) at (1.5,1) [D5] {};
                \node (g4) at (1,1) [D5] {};
                \draw [dotted_brane] (-2,1) -- (-1.5,1);
                \draw [dashed_brane] (-1.5,1) -- (-1,1);
                \draw [dotted_brane] (-1,1) -- (-0.5,1);
                \draw [dashed_brane] (0.5,1) -- (1,1);
                \draw [dotted_brane] (1,1) -- (1.5,1);
                \draw [dashed_brane] (1.5,1) -- (2,1);
                \node at (0,2) {$\ell$};
                \node at (0, 1) [scale=1]{$\cdots$} {};
                \draw [thick,decoration={brace,mirror,raise=0.25cm},decorate] (-1.75,1) -- (1.75,1);
                \node at (0,0.5) {$2n+1$};
            \end{tikzpicture}$};
            \node (b) at (4,0) {$\begin{tikzpicture}
                \node at (0,0) [scale=1]{$\longleftrightarrow$};
            \end{tikzpicture}$};
            \node (c) at (6,0) {$\begin{tikzpicture}
                \node (1) [gauge, label=below:{$\mathfrak{c}_{\ell}$}] at (0,0) {};
                \node (2) [flavor, label=above:{$\mathfrak{b}_{n}$}] at (0,1) {};
                \draw[-] (1)--(2);
            \end{tikzpicture}$};
        \end{tikzpicture}
        \label{rule_C_gauge_B_flavour}
    \end{equation}
        \begin{equation}
        \centering
        \begin{tikzpicture}
            \node (a) at (0,0) {$\begin{tikzpicture}
                \draw[-] (-2,0) -- (-2,2);
                \draw[-] (2,0) -- (2,2);
                \draw[-] (-2,1.75) -- (2,1.75);
                \draw[-] (-2,0.25) -- (2,0.25);
                \node (g1) at (-1.5,1) [D5] {};
                \node (g2) at (-1,1) [D5] {};
                \node (g3) at (1.5,1) [D5] {};
                \node (g4) at (1,1) [D5] {};
                \draw [dashed_brane] (-2,1) -- (-1.5,1);
                \draw [dotted_brane] (-1.5,1) -- (-1,1);
                \draw [dashed_brane] (-1,1) -- (-0.5,1);
                \draw [dashed_brane] (0.5,1) -- (1,1);
                \draw [dotted_brane] (1,1) -- (1.5,1);
                \draw [dashed_brane] (1.5,1) -- (2,1);
                \node at (0,2) {$\ell$};
                \node at (0, 1) [scale=1]{$\cdots$} {};
                \draw [thick,decoration={brace,mirror,raise=0.25cm},decorate] (-1.75,1) -- (1.75,1);
                \node at (0,0.5) {$2n$};
            \end{tikzpicture}$};
            \node (b) at (4,0) {$\begin{tikzpicture}
                \node at (0,0) [scale=1]{$\longleftrightarrow$};
            \end{tikzpicture}$};
            \node (c) at (6,0) {$\begin{tikzpicture}
                \node (1) [gauge, label=below:{$\mathfrak{c}_{\ell}$}] at (0,0) {};
                \node (2) [flavor, label=above:{$\mathfrak{d}_{n}$}] at (0,1) {};
                \draw[-] (1)--(2);
            \end{tikzpicture}$};
        \end{tikzpicture}
        \label{rule_C_gauge_D_flavour}
    \end{equation}
    \begin{equation}
        \centering
        \begin{tikzpicture}
            \node (a) at (0,0) {$\begin{tikzpicture}
                \draw[-] (-2,0) -- (-2,2);
                \draw[-] (2,0) -- (2,2);
                \draw[-] (-2,1.75) -- (2,1.75);
                \draw[-] (-2,0.25) -- (2,0.25);
                \node (g1) at (-1.5,1) [D5] {};
                \node (g2) at (-1,1) [D5] {};
                \node (g3) at (1.5,1) [D5] {};
                \node (g4) at (1,1) [D5] {};
                \draw[-] (-2,1) -- (-1.5,1);
                \draw[-] (-1,1) -- (-0.5,1);
                \draw[-] (0.5,1) -- (1,1);
                \draw[-] (1.5,1) -- (2,1);
                \node at (0,2) {$\ell$};
                \node at (0, 1) [scale=1]{$\cdots$} {};
                \draw [thick,decoration={brace,mirror,raise=0.25cm},decorate] (-1.75,1) -- (1.75,1);
                \node at (0,0.5) {$2n$};
                \draw[-] (-1.38,1.12)--(-1.12,1.12);
                \draw[-] (-1.38,0.88)--(-1.12,0.88);
                \draw[-] (1.38,1.12)--(1.12,1.12);
                \draw[-] (1.38,0.88)--(1.12,0.88);
                \draw [dotted_brane] (-2,1) -- (-2.5,1);
                \draw [dotted_brane] (2.5,1) -- (2,1);
            \end{tikzpicture}$};
            \node (b) at (4,0) {$\begin{tikzpicture}
                \node at (0,0) [scale=1]{$\longleftrightarrow$};
            \end{tikzpicture}$};
            \node (c) at (6,0) {$\begin{tikzpicture}
                \node (1) [gauge, label=below:{$\mathfrak{b}_{\ell}$}] at (0,0) {};
                \node (2) [flavor, label=above:{$\mathfrak{c}_{n}$}] at (0,1) {};
                \draw[-] (1)--(2);
            \end{tikzpicture}$};
        \end{tikzpicture}
        \label{rule_B_gauge_C_flavour}
    \end{equation}
    \item As has already been mentioned, the interpretation of $D$-type gauge nodes is more subtle than the $A-/B-/C-$ cases. Modulo $\orm/\sorm$ ambiguities, $\sorm(2n)$ gauge groups are read from brane systems as shown in \eqref{rule:SO_type_gauge}. Section \ref{sec:Rule_OvsSO} will consider this in further detail.
    \begin{equation}
        \centering
        \begin{tikzpicture}
            \node (a) at (0,0) {$\begin{tikzpicture}
                \draw[-] (-1,0) -- (-1,2);
                \draw[-] (1,0) -- (1,2);
                \draw[-] (-1,1.5) -- (1,1.5);
                \draw[-] (-1,0.5) -- (1,0.5);
                \draw[-] (1,0.25) -- (1.75,0.25);
                \draw[-] (-1,0.25) -- (-1.75,0.25);
                \draw[-] (1,1.75) -- (1.75,1.75);
                \draw[-] (-1,1.75) -- (-1.75,1.75);
                \draw [dashed_brane] (1,1) -- (1.75,1);
                \draw [dashed_brane] (-1,1) -- (-1.75,1);
                \node at (0,1.8) {$\ell$};
                \node at (1.5,2) {$m$};
                \node at (-1.5,2) {$n$};
            \end{tikzpicture}$};
            \node (b) at (4,0) {$\begin{tikzpicture}
                \node at (0,0) [scale=1]{$\longleftrightarrow$};
            \end{tikzpicture}$};
            \node (c) at (8,0) {$\begin{tikzpicture}
            \node (1) [gauge, label=below:{$\mathfrak{d}_{\ell}$}] at (0,0) {};
            \node (3) [gauge, label=below:{$\mathfrak{c}_n$}] at (-1.5,0) {};
            \node (4) [gauge, label=below:{$\mathfrak{c}_m$}] at (1.5,0) {};
            \node (6) at (-2.5, 0) [scale=1]{$\cdots$} {};
            \node (7) at (2.5, 0) [scale=1]{$\cdots$} {};
            \draw[-] (6) -- (3) -- (1) -- (4) -- (7);
            \end{tikzpicture}$};
        \end{tikzpicture}
        \label{rule:SO_type_gauge}
    \end{equation}
\end{itemize}
\paragraph{Semi-Infinite $\widetilde{O3}^{-}$ Planes}
A further condition exists for brane systems with semi-infinite $\widetilde{O3}^{-}$ planes of the sort seen in \Figref{fig:semi_inf_O3minustilde_plane}. The na\"ive quiver derivation would result in a theory with a Witten anomaly \cite{WITTEN1982324} from the $C$-type gauge node seeing an odd number of half-hypers. 

To amend this, it is useful to imagine the $\widetilde{O3}^{-}$ plane ending at infinity on a $\frac{1}{2}$D5-brane. This can then be brought in and HW-transitioned as shown in \Figref{fig:semi_inf_O3minustilde_plane}, removing the $\widetilde{O3}^{-}$ plane from the brane system and resulting in a non-anomalous quiver \cite{Feng_2000}. Note that this HW-transition is the only such allowed that neither creates nor annihilates D3-branes --- put simply, the trick given in \Figref{fig:semi_inf_O3minustilde_plane} would not work for semi-infinite $O3$ planes of any other type.
\subsection{$\orm(2n)$ versus $\sorm(2n)$}
\label{sec:Rule_OvsSO}
As mentioned in the previous section, reading the global form of the gauge group for gauge nodes with $D$-type algebras is unreliable. Appealing to past examples in \cite{Cabrera:2017njm}, as well as a further simple example considered here, this section conjectures that the correct global form of a gauge node with $D$-type algebra is controlled by the presence of $\frac{1}{2}$D5-branes on the $O3^{+}$ planes either side of the interval. The prescription is as follows.
\begin{itemize}
    \item If there are zero D5-branes on either side of the interval, the gauge group $\sorm(2\ell)$ is read from $\ell$ D3-branes suspended between the two $\frac{1}{2}$NS5-branes as in \eqref{rule:SO_type_gauge}. 
    \item If a single $\frac{1}{2}$D5-brane is present on either side of the interval, the gauge group $O(2\ell)$ is read and the two $C$-type gauge nodes on either side pick up $\mathfrak{b}_{0}$ flavours as in \eqref{rule:O_type_gauge}. Multiple $\frac{1}{2}$D5-branes on the $O3^{+}$ planes on either side of the $O3^{-}$ interval will also give rise to an $O(2\ell)$ gauge node, with flavours on the two $C$-type gauge nodes consistent with \eqref{rule_C_gauge_B_flavour} and \eqref{rule_C_gauge_D_flavour}. 
    \begin{equation}
        \centering
        \begin{tikzpicture}
            \node (a) at (0,0) {$\begin{tikzpicture}
            \draw[-] (-1,0) -- (-1,2);
            \draw[-] (1,0) -- (1,2);
            \node (g1) at (-1.5,1) [D5] {};
            \node (g2) at (1.5,1) [D5] {};
            \draw[-] (-1,1.5) -- (1,1.5);
            \draw[-] (-1,0.5) -- (1,0.5);
            \draw[-] (1,0.25) -- (1.75,0.25);
            \draw[-] (-1,0.25) -- (-1.75,0.25);
            \draw[-] (1,1.75) -- (1.75,1.75);
            \draw[-] (-1,1.75) -- (-1.75,1.75);
            \draw [dashed_brane] (1,1) -- (1.5,1);
            \draw [dotted_brane] (1.5,1) -- (1.75,1);
            \draw [dashed_brane] (-1,1) -- (-1.5,1);
            \draw [dotted_brane] (-1.5,1) -- (-1.75,1);
            \node at (0,1.8) {$\ell$};
            \node at (1.5,2) {$m$};
            \node at (-1.5,2) {$n$};
            \end{tikzpicture}$};
            \node (b) at (4,0) {$\begin{tikzpicture}
                \node at (0,0) [scale=1]{$\longleftrightarrow$};
            \end{tikzpicture}$};
            \node (c) at (8,0) {$\begin{tikzpicture}
            \node (1) [gauge, label=below:{$\orm(2\ell)$}] at (0,0) {};
            \node (2) [flavor, label=above:{$\mathfrak{b}_{0}$}] at (-1.5,1) {};
            \node (3) [gauge, label=below:{$\mathfrak{c}_n$}] at (-1.5,0) {};
            \node (4) [gauge, label=below:{$\mathfrak{c}_m$}] at (1.5,0) {};
            \node (5) [flavor, label=above:{$\mathfrak{b}_{0}$}] at (1.5,1) {};
            \node (6) at (-2.5, 0) [scale=1]{$\cdots$} {};
            \node (7) at (2.5, 0) [scale=1]{$\cdots$} {};
            \draw[-] (6) -- (3) -- (2) -- (3) -- (1) -- (4) -- (5) -- (4) -- (7);
            \end{tikzpicture}$};
        \end{tikzpicture}
        \label{rule:O_type_gauge}
    \end{equation}
\end{itemize}
\paragraph{Example 1: Next-to-Minimal $\sorm(2n+1)$}
One of the most straightforward examples concerns the quiver \eqref{quiv:nminBQuiv}, derived from the brane system in \Figref{fig:Ex1:nminB}, whose Coulomb branch is the closure of the next-to-minimal orbit of $\sorm(2k+1)$, $\overline{O}^{B_k}_{(3,1^{2k-2})}$ \cite{Cabrera:2017njm}. 
\begin{figure}[h]
    \centering
    \begin{tikzpicture}
        \draw[dotted_brane] (-1,1) -- (-0.5,1);
        \node (g1) at (-0.5,1) [D5] {};
        \draw[dashed_brane] (-0.5,1) -- (0,1);
        \draw[-](0,0)--(0,2);
        \draw[-](1,0)--(1,2);
        \node (g2) at (1.5,1) [D5] {};
        \draw[-](2,0)--(2,2);
        \draw[-](3,0)--(3,2);
        \draw[dashed_brane] (1,1)--(1.5,1);
        \draw[dotted_brane] (1.5,1)--(2,1);
        \draw[-] (2,1)--(3,1);
        \draw[-](0,1.5)--(1,1.5);
        \draw[-](0,0.5)--(1,0.5);
        \draw[-](1,1.75)--(2,1.75);
        \draw[-](1,0.25)--(2,0.25);
        \draw[-](2,1.5)--(3,1.5);
        \draw[-](2,0.5)--(3,0.5);
        \draw[-](3,1.75)--(3.5,1.75);
        \draw[-](3,0.25)--(3.5,0.25);
        \draw[dotted_brane] (3,1)--(3.5,1);
        \node at (4.5, 0.97) [scale=2]{$\cdots$} {};
        \draw[dotted_brane] (5.5,1)--(6,1);
        \draw[-](5.5,1.75)--(6,1.75);
        \draw[-](5.5,0.25)--(6,0.25);
        \draw[-](7,1.5)--(6,1.5);
        \draw[-](7,0.5)--(6,0.5);
        \draw[-](8,1.75)--(7,1.75);
        \draw[-](8,0.25)--(7,0.25);
        \draw[-](9,1.5)--(8,1.5);
        \draw[-](9,0.5)--(8,0.5);
        \draw[dotted_brane] (7,1)--(7.5,1);
        \draw[dashed_brane] (7.5,1)--(8,1);
        \draw[-](6,0)--(6,2);
        \draw[-](6,1)--(7,1);
        \draw[-](7,0)--(7,2);
        \node (g2) at (7.5,1) [D5] {};
        \draw[-](9,0)--(9,2);
        \draw[-] (8,0)--(8,2);
    \end{tikzpicture}
    \caption{The brane system giving rise to the quiver \eqref{quiv:nminBQuiv}, whose Coulomb branch is $\overline{O}^{B_k}_{(3,1^{2k-2})}$ and whose Higgs branch is $D_{k+1}$ \cite{Cabrera:2017njm}. The presence of the $\frac{1}{2}$D5-brane on the far left is conjectured to gauge a $\mathbb{Z}_{2}$ resulting in the leftmost $D$-type gauge node of \eqref{quiv:nminBQuiv} becoming $\orm(2)$.}
    \label{fig:Ex1:nminB}
\end{figure}
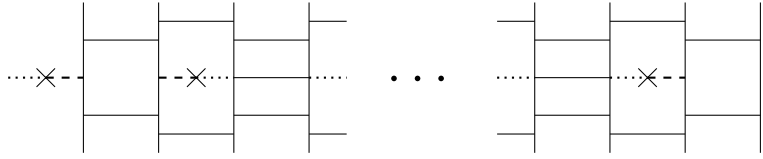
\begin{equation}
\begin{tikzpicture}
    \node (10) [gauge, label=below:{$\orm(2)$}] at (-1,0) {};
    \node (1) [gauge, label=below:{$\mathfrak{c}_1$}] at (0,0) {};
    \node (2) [gauge, label=below:{$\mathfrak{b}_1$}] at (1,0) {};
    \node (3) [gauge, label=below:{$\mathfrak{b}_1$}] at (4,0) {};
    \node (4) [gauge, label=below:{$\mathfrak{c}_1$}] at (5,0) {};
    \node (5) [flavor, label=above:{$\mathfrak{b}_0$}] at (0,1) {};
    \node (6) [flavor, label=above:{$\mathfrak{b}_0$}] at (5,1) {};
    \node (7) [gauge, label=below:{$\mathfrak{d}_1$}] at (6,0) {};
    \draw (10)--(1)--(5)--(1)--(2)--(1.75,0);
    \draw (3.25,0)--(3)--(4)--(6)--(4)--(7);
    \node at (2.54, -0.03) [scale=2]{$\cdots$} {};
    \draw [thick,decoration={brace,mirror,raise=0.75cm},decorate] (-0.25,0) -- (5.25,0);
    \node at (2.5,-1.25) {$2k-3$};
\end{tikzpicture}
\label{quiv:nminBQuiv}
\end{equation}
\begin{figure}[h!]
    \centering
    \begin{subfigure}{0.49\textwidth}
    \centering
    \begin{tikzpicture}
        \draw[-](1,1.75)--(4.25,1.75);
        \draw[-](1,0.25)--(4.25,0.25);
        \draw[dotted_brane](-0.5,1)--(0,1);
        \draw[-](0,0)--(0,2);
        \draw[-](0,1)--(1,1);
        \draw[-](1,0)--(1,2);
        \draw[dotted_brane](1,1)--(1.5,1);
        \node (g1) at (1.5,1) [D5] {};
        \draw[dashed_brane](1.5,1)--(2,1);
        \node at (2.75, 0.97) [scale=2]{$\cdots$} {};
        \draw[dotted_brane](3.25,1)--(3.75,1);
        \node (g1) at (3.75,1) [D5] {};
        \draw[dashed_brane](3.75,1)--(4.25,1);
        \draw[-](4.25,0)--(4.25,2);
        \draw [thick,decoration={brace,mirror,raise=0.25cm},decorate] (1.25,1) -- (4,1);
        \node at (2.5,0.5) {$2k+1$};
    \end{tikzpicture}
    \caption{}
    \label{fig:b+qOneZ2Gauged1}
    \end{subfigure}
    \begin{subfigure}{0.49\textwidth}
    \centering
    \begin{tikzpicture}
        \node (1) [gauge, label=below:{$\mathfrak{b}_{0}$}] at (0,0) {};
        \node (2) [gauge, label=below:{$\mathfrak{c}_{1}$}] at (1,0) {};
        \node (3) [flavour, label=above:{$\mathfrak{b}_{k}$}] at (1,1) {};
        \draw[-] (1)--(2)--(3);
    \end{tikzpicture}
    \caption{}
    \label{fig:b+qOneZ2Gauged2}
    \end{subfigure}
    \caption{The brane system in \Figref{fig:b+qOneZ2Gauged1}, reached after performing $S$-duality on that given in \Figref{fig:Ex1:nminB}, gives rise to the quiver in \Figref{fig:b+qOneZ2Gauged2}.}
    \label{fig:b+qOneZ2Gauged}
\end{figure}
The Higgs and Coulomb branches of the corresponding 3d mirror, given in \Figref{fig:b+qOneZ2Gauged2} and derived from the brane system in \Figref{fig:b+qOneZ2Gauged1}, are $\overline{O}^{B_k}_{(3,1^{2k-2})}$ and $D_{k+1}$ (which is the top slice in the nilcone of $\mathfrak{c}_{k}$ \cite{Grimminger:2020dmg}) respectively. Without the $\mathbb{Z}_{2}$ gauging sending $\sorm(2)$ to $\orm(2)$ in \eqref{quiv:nminBQuiv}, the moduli spaces of \eqref{quiv:nminBQuiv} would not match those of the dual \eqref{fig:b+qOneZ2Gauged2} derived from the brane system. This is one of the simplest cases wherein the $\mathbb{Z}_{2}$ quotient plays a crucial role in determining the resulting moduli spaces.
\paragraph{Example 2:} Consider the brane system given in \Figref{fig:branes_example_2i_A}. Using the rule in \eqref{rule:O_type_gauge}, the gauge node with $D$-type gauge algebra should be read as $\orm(2)$ instead of $\sorm(2)$ -- both possibilities are given in \Figref{fig:branes_example_2i_B}. Performing S-duality on the brane system to derive the 3d mirror theory results in the configuration given in \Figref{fig:branes_example_2ii_A}, which unambiguously gives the corresponding quiver \Figref{fig:branes_example_2ii_B}. The Coulomb branch of \Figref{fig:branes_example_2ii_B} is insensitive to the $\mathfrak{b}_{0}$ gauging and in \Figref{fig:branes_example_2i_A} both $G=\sorm(2)$ and $G=\orm(2)$ give the same Higgs branch \cite{Hanany:2016gbz}. Hence it is the Coulomb branch of \Figref{fig:branes_example_2i_B} and the Higgs branch of \Figref{fig:branes_example_2ii_B} which provide the nontrivial check. The Hilbert series for the Higgs branch of \Figref{fig:branes_example_2ii_B} is given in \eqref{eqn:HS_OvsSO_Ex2_A}. Upon computing the Coulomb branches of the two candidate quivers in \Figref{fig:branes_example_2i_B}, only the quiver with the $\orm(2)$ gauge node gives the same (unrefined) Hilbert series as \eqref{eqn:HS_OvsSO_Ex2_A}, shown in \eqref{eqn:HS_OvsSO_Ex2_B}. In the brane system \Figref{fig:branes_example_2i_A}, this agrees with the presence of $\frac{1}{2}$D5-branes either side of the leftmost $\frac{1}{2}$NS5-brane interval. 
\begin{equation}
    \text{HS}_{\mathcal{H}}(\mathcal{Q}_{\ref{fig:branes_example_2ii_B}}) =\begin{aligned} &\pe\left[([4]-[2])t^{4}\right]\\&\times(1+[2]t^2+(2[2]+[1])t^{4}+([4]+[2]+2[1])t^{6}+\cdots+t^{12})\end{aligned}
    \label{eqn:HS_OvsSO_Ex2_A}
    \end{equation}
    \begin{equation}
    \text{HS}_{\mathcal{H}}(\mathcal{Q}_{\ref{fig:branes_example_2ii_B}})\rvert_{a,b\rightarrow1} = \frac{1+t^{2}+4t^4+t^6+t^{8}}{(1-t^2)^{2}(1-t^4)^{2}} = \text{HS}_{\mathcal{C}}(\mathcal{Q}_{\ref{fig:branes_example_2i_A}}\rvert_{G = \orm(2)})
\label{eqn:HS_OvsSO_Ex2_B}
\end{equation}
\begin{figure}[h!]
    \centering
    \begin{subfigure}{0.49\textwidth}
    \centering
    \begin{tikzpicture}
        \draw[-] (0,0) -- (0,2);
        \draw[-] (2,0)--(2,2);
        \draw[-] (4,0)--(4,2);
        \draw[dotted_brane] (-1,1)--(-0.5,1); 
        \draw[dashed_brane] (-0.5,1)--(0,1); 
        \node (g1) at (-0.5,1) [D5] {};
        \node (g2) at (2.4,1) [D5] {};
        \node (g3) at (2.8,1) [D5] {};
        \node (g4) at (3.2,1) [D5] {};
        \node (g5) at (3.6,1) [D5] {};
        \draw[dashed_brane] (2,1)--(2.4,1); 
        \draw[dotted_brane] (2.4,1)--(2.8,1);
        \draw[dashed_brane] (2.8,1)--(3.2,1);
        \draw[dotted_brane] (3.2,1)--(3.6,1);
        \draw[dashed_brane] (3.6,1)--(4,1);
        \draw[-] (0,1.8)--(2,1.8);
        \draw[-] (0,0.2)--(2,0.2);
        \draw[-] (2,0.6)--(4,0.6);
        \draw[-] (2,1.4)--(4,1.4);
    \end{tikzpicture}
    \caption{}
    \label{fig:branes_example_2i_A}
    \end{subfigure}
    \begin{subfigure}{0.49\textwidth}
    \centering
    \begin{tikzpicture}
    \node (1) [gauge, label=below:{$G$}] at (0,0) {};
    \node (2) [gauge, label=below:{$\mathfrak{c}_{1}$}] at (1,0) {};
    \node (3) [flavor, label=above:{$\mathfrak{d}_2$}] at (1,1) {};
    \draw (1)--(2)--(3);
    \end{tikzpicture}
    \caption{}
    \label{fig:branes_example_2i_B}
    \end{subfigure}
    \begin{subfigure}{0.49\textwidth}
    \centering
    \begin{tikzpicture}
    \draw[dotted_brane] (-1.5,1)--(-1,1); 
    \draw[-] (-1,0) -- (-1,2);
    \draw[-] (0,0) -- (0,2);
    \draw[-] (2,0)--(2,2);
    \draw[-] (3,0)--(3,2);
    \draw[-] (4,0)--(4,2);
    \draw[-] (-1,1) -- (0,1);
    \draw[dotted_brane] (0,1)--(0.5,1); 
    \draw[dashed_brane] (0.5,1)--(1,1); 
    \draw[dotted_brane] (1,1)--(1.5,1); 
    \draw[dashed_brane] (1.5,1)--(2,1); 
    \node (g1) at (0.5,1) [D5] {};
    \node (g2) at (1,1) [D5] {};
    \node (g3) at (1.5,1) [D5] {};
    \draw[-] (0,1.8)--(2,1.8);
    \draw[-] (0,0.2)--(2,0.2);
    \draw[-] (2,0.6)--(3,0.6);
    \draw[-] (2,1.4)--(3,1.4);
    \draw[dashed_brane] (3,1)--(4,1); 
    \end{tikzpicture}
    \caption{}
    \label{fig:branes_example_2ii_A}
    \end{subfigure}
    \begin{subfigure}{0.49\textwidth}
    \centering
    \begin{tikzpicture}
    \node (1) [gauge, label=below:{$\mathfrak{b}_0$}] at (0,0) {};
    \node (2) [gauge, label=below:{$\mathfrak{c}_{1}$}] at (1,0) {};
    \node (3) [gauge, label=below:{$\mathfrak{d}_{1}$}] at (2,0) {};
    \node (4) [flavor, label=above:{$\mathfrak{b}_1$}] at (1,1) {};
    \draw (1)--(2)--(4)--(2)--(3);
    \end{tikzpicture}
    \caption{}
    \label{fig:branes_example_2ii_B}
    \end{subfigure}
    \caption{The leftmost $\frac{1}{2}$NS5-brane interval in the brane system in \Figref{fig:branes_example_2i_A} can be argued to give rise to either a $G = \sorm(2)$ or a $G = \orm(2)$ gauge node. It turns out that only $G = \orm(2)$ is the consistent choice. \Figref{fig:branes_example_2ii_A} gives the brane system obtained from \Figref{fig:branes_example_2i_A} after S-duality, the resulting theory is given in \Figref{fig:branes_example_2ii_B}.}
    \label{}
\end{figure}
\paragraph{Lusztig's Canonical Quotient}
Many examples of the phenomenon considered in \eqref{rule:O_type_gauge} can be found in \cite{Cabrera:2017ucb}. In these cases, the presence of a $\mathbb{Z}_{2}$ gauging implied the existence of an $\bar{A}(\mathcal{O})=\mathbb{Z}_{2}$ canonical quotient \cite{Generic_singularities,SOMMERS2001790, achar_lcq} on the Coulomb branch. Of course, given that \eqref{rule:O_type_gauge} ascribes a $\mathbb{Z}_{2}$ quotient to a relatively arbitrary brane system -- crucially, one whose associated Coulomb branch is not necessarily a nilpotent orbit closure -- the question arises as to the interpretation of this finite group data \emph{outside} the nilcone.

\begin{equation}
    \begin{tikzpicture}
        \node (1) [gauge, label=below:{$\mathfrak{d}_1$}] at (0,0) {};
        \node (2) [gauge, label=below:{$\mathfrak{c}_1$}] at (1,0) {};
        \node (3) [gauge, label=below:{$\mathfrak{d}_2$}] at (2,0) {};
        \node (4) [gauge, label=below:{$\mathfrak{c}_2$}] at (3,0) {};
        \node (5) [gauge, label=below:{$\mathfrak{d}_3$}] at (4,0) {};
        \node (6) [gauge, label=below:{$\mathfrak{c}_3$}] at (5,0) {};
        \node (7) [gauge, label=below:{$\mathfrak{d}_4$}] at (6,0) {};
        \node (8) [gauge, label=below:{$\mathfrak{c}_4$}] at (7,0) {};
        \node (9) [gauge, label=below:{$\orm(8)$}] at (8,0) {};
        \node (10) [gauge, label=below:{$\mathfrak{c}_3$}] at (9,0) {};
        \node (11) [gauge, label=below:{$\orm(4)$}] at (10,0) {};
        \node (12) [flavor, label=above:{$\mathfrak{d}_1$}] at (7,1) {};
        \node (13) [flavor, label=above:{$\mathfrak{d}_1$}] at (9,1) {};
        \draw[-] (1) -- (2) --(3) -- (4) -- (5) -- (6) --(7) -- (8) -- (9) -- (10) -- (11);
        \draw[-] (12) -- (8);
        \draw[-] (13) -- (10);
    \end{tikzpicture}
    \label{LCQ_example}
\end{equation}
Following \cite{Cabrera:2017ucb}, the canonical quotient of Lusztig (LCQ) associated to a nilpotent orbit closure is read from a framed orthosymplectic quiver by considering the number of $\mathbb{Z}_{2}$ gaugings applied to nodes with $D$-type algebras. For instance, the $T_{[4^{2},2^{2}]}(\sprm(6))$ theory in \eqref{LCQ_example}, whose Coulomb branch is $\bar{\mathcal{O}}^{[5,3^{2},1^{2}]}$ in the nilpotent cone of $\mathfrak{so}_{13}$ \cite{Cabrera:2017ucb} has $\bar{A}(\mathcal{O}) = (\mathbb{Z}_{2})^{2}$ -- one $\mathbb{Z}_{2}$ from each gauge node of the form $\orm(2\ell)$. Diagonally-gauged $\sorm$-nodes -- to be considered further in Section \ref{sec:gauging_elec_magnet_perspec} -- similarly contribute a single $\mathbb{Z}_{2}$. In the language of \cite{Cabrera:2017ucb}, the LCQ of a quiver whose Coulomb branch is a nilpotent orbit closure is $(\mathbb{Z}_{2})^{m}$, where $m$ is the number of `chains' in the quiver. 

Although the LCQ is read differently in the case of unitary quivers to the rule given here, the generality of these quiver constructions leads to the question of extending the definition of the LCQ outside nilpotent cones. It is tempting to conjecture that the Coulomb branch of any framed orthosymplectic quiver with gauged $\mathbb{Z}_{2}$ factors induced by chains has some analogue of an LCQ given as $(\mathbb{Z}_{2})^{k}$, where $k$ is the number of chains. \footnote{Similarly, it can be conjectured that the Coulomb branch of any decorated unframed quiver (unitary or orthosymplectic) has an associated $S_n$ `Lusztig canonical quotient' corresponding to the number of decorated gauge nodes of rank 1 \cite{Bourget:2022tmw, Bourget:2022ehw,Hanany:2018vph}.} Moreover, it would be interesting to consider if such an argument could be extended to decorations of arbitrary sub-quivers giving isolated singularities, making the connection with gauging finite symmetries clearer.
\section{A Motivating Example: \boldmath$\orm(3)$ with $N_f$ Flavours\unboldmath}
\label{sec:motivating_example_O3_Nf}
One of the simplest examples of the manipulations studied in this work concerns $\orm(3)$ gauge theory with $N_f$ flavours rotated under $\sprm(N_f)$. The Higgs branch is well known to be the closure of the nilpotent orbit $[2^{3},1^{2n-6}]$, with Hasse diagram as given in \Figref{hasse:O3_Nf}. In a brane system, the $\orm(3)$ gauge theory can be realised as in \Figref{fig:branes_O3_higgsing_A}, where the $\frac{1}{2}$D5-branes are split along the $O3$ orientifolds. Higgsing the theory corresponds to performing a Kraft-Procesi transition on the $\frac{1}{2}$D3-branes created between $\frac{1}{2}$D5-branes on $\widetilde{O3^{-}}$ planes, highlighted in red in \Figref{fig:branes_O3_higgsing_B}; after X-collapse, the brane system is as shown in \Figref{fig:branes_O3_higgsing_C} -- naively, this would be interpreted as an $\sorm(2)$ gauge theory with $N_f-1$ flavours. However, in accordance with the rule given in Section \eqref{rule:O_type_gauge}, the gauge group is instead read as $\orm(2)$. This agrees with the field theory, which stipulates that the Higgsing returns an $\orm(2)$ gauge theory.
\begin{figure}[h!]
\centering
\begin{tikzpicture}
    \node at (1,-6.2) {$\begin{tikzpicture}
        \node (1) [gauge, label=below:{$\orm(3)$}] at (0,0) {};
        \node (2) [flavor, label=below:{$\mathfrak{c}_{N_f}$}] at (1,0) {};
        \draw[-] (1)--(2);
    \end{tikzpicture}$};
    \node at (1,-4.2) {$\begin{tikzpicture}
        \node (1) [gauge, label=below:{$\orm(2)$}] at (0,0) {};
        \node (2) [flavor, label=below:{$\mathfrak{c}_{N_f-1}$}] at (1,0) {};
        \draw[-] (1)--(2);
    \end{tikzpicture}$};
    \node at (1,-2.2) {$\begin{tikzpicture}
        \node (1) [gauge, label=below:{$\orm(1)$}] at (0,0) {};
        \node (2) [flavor, label=below:{$\mathfrak{c}_{N_f-2}$}] at (1,0) {};
        \draw[-] (1)--(2);
    \end{tikzpicture}$};
    \node (1) [hasse, label=left:$3N_f-3$] at (0,0) {};
    \node (2) [hasse, label=left:$2N_f-1$] at (0,-2) {};
    \node (3) [hasse, label=left:$N_f$] at (0,-4) {};
    \node (4) [hasse, label=left:$0$] at (0,-6) {};
    \draw[-] (4) -- (3) node[pos=0.5,midway, left]{$c_{N_f}$} -- (2) node[pos=0.5,midway, left]{$c_{N_f-1}$} -- (1) node[pos=0.5,midway, left]{$c_{N_f-2}$};
\end{tikzpicture}
\caption{The Hasse diagram for the Higgs branch of $\orm(3)$ gauge theory with $N_f$ flavours, with leaves labelled by their dimension alongside the Higgsing pattern down to $\orm(1)$ gauge theory with $N_f-2$ flavours.}
\label{hasse:O3_Nf}
\end{figure}
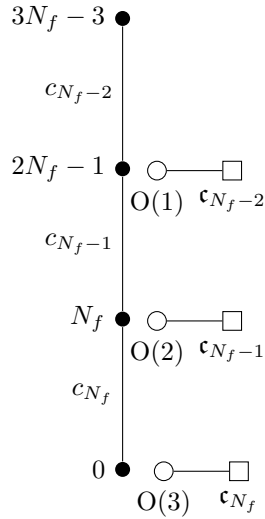

\begin{figure}[h!]
\centering
\begin{subfigure}{0.49\textwidth}
\centering
\begin{tikzpicture}
    \draw[dotted_brane] (-2.5,1)--(-2,1);
    \draw[dotted_brane] (2.5,1)--(2,1);
    \draw[-] (-2,0) -- (-2,2);
    \draw[-] (2,0) -- (2,2);
    \draw[-] (-2,1.75) -- (2,1.75);
    \draw[-] (-2,0.25) -- (2,0.25);
    \node (g1) at (-1.5,1) [D5] {};
    \node (g2) at (-1,1) [D5] {};
    \node (g3) at (1.5,1) [D5] {};
    \node (g4) at (1,1) [D5] {};
    \draw[-] (-2,1) -- (-1.5,1);
    \draw[-] (-1,1) -- (-0.5,1);
    \draw[-] (0.5,1) -- (1,1);
    \draw[-] (1.5,1) -- (2,1);
    \node at (0, 1) [scale=1]{$\cdots$} {};
    \draw [thick,decoration={brace,mirror,raise=0.25cm},decorate] (-1.75,1) -- (1.75,1);
    \node at (0,0.45) {$2N_f$};
    \draw[-] (-1.38,1.12)--(-1.12,1.12);
    \draw[-] (-1.38,0.88)--(-1.12,0.88);
    \draw[-] (1.38,1.12)--(1.12,1.12);
    \draw[-] (1.38,0.88)--(1.12,0.88);
\end{tikzpicture}
\caption{}
\label{fig:branes_O3_higgsing_A}
\end{subfigure}
\begin{subfigure}{0.49\textwidth}
\centering
\begin{tikzpicture}
    \draw[dotted_brane] (-2.5,1)--(-2,1);
    \draw[dotted_brane] (2.5,1)--(2,1);
    \draw[-] (-2,0) -- (-2,2);
    \draw[-] (2,0) -- (2,2);
    \draw[-] (-2,1.75) -- (2,1.75);
    \draw[-] (-2,0.25) -- (2,0.25);
    \node (g1) at (-1.5,1) [D5] {};
    \node (g2) at (-1,1) [D5] {};
    \node (g3) at (1.5,1) [D5] {};
    \node (g4) at (1,1) [D5] {};
    \draw[-] (-2,1) -- (-1.5,1);
    \draw[-] (-1,1) -- (-0.5,1);
    \draw[-] (0.5,1) -- (1,1);
    \draw[-] (1.5,1) -- (2,1);
    \node at (0, 1) [scale=1]{$\cdots$} {};
    \draw [thick,decoration={brace,mirror,raise=0.25cm},decorate] (-1.75,1) -- (1.75,1);
    \node at (0,0.45) {$2N_f$};
    \draw[-,color=red] (-1.38,1.12)--(-1.12,1.12);
    \draw[-,color=red] (-1.38,0.88)--(-1.12,0.88);
    \draw[-,color=red] (1.38,1.12)--(1.12,1.12);
    \draw[-,color=red] (1.38,0.88)--(1.12,0.88);
\end{tikzpicture}
\caption{}
\label{fig:branes_O3_higgsing_B}
\end{subfigure}
\begin{subfigure}{0.49\textwidth}
\centering
\begin{tikzpicture}
    \draw[dotted_brane] (-3,1)--(-2.5,1);
    \draw[dashed_brane] (-2.5,1)--(-2,1);
    \draw[dashed_brane] (2,1)--(2.5,1);
    \draw[dotted_brane] (2.5,1)--(3,1);
    \draw[-] (-2,0) -- (-2,2);
    \draw[-] (2,0) -- (2,2);
    \draw[-] (-2,1.75) -- (2,1.75);
    \draw[-] (-2,0.25) -- (2,0.25);
    \node (g1) at (-2.5,1) [D5] {};
    \node (g2) at (-1.5,1) [D5] {};
    \node (g3) at (2.5,1) [D5] {};
    \node (g4) at (1.5,1) [D5] {};
    \draw[-] (-1.5,1) -- (-1,1);
    \draw[-] (1,1) -- (1.5,1);
    \node at (0, 1) [scale=1]{$\cdots$} {};
    \draw [thick,decoration={brace,mirror,raise=0.25cm},decorate] (-1.75,1) -- (1.75,1);
    \node at (0,0.45) {$2N_f-2$};
\end{tikzpicture}
\caption{}
\label{fig:branes_O3_higgsing_C}
\end{subfigure}
\caption{The Higgsing process taking the $\orm(3)$ gauge theory with $N_f$ flavours, read from the brane system in \Figref{fig:branes_O3_higgsing_A}, to the $\orm(2)$ gauge theory with $N_f-1$ flavours. The $\frac{1}{2}$D3-branes created under the splitting of a D5-brane on an $\widetilde{O3^{-}}$ plane, represented in red in \Figref{fig:branes_O3_higgsing_B}, undergo a Kraft-Procesi transition resulting in the brane system in \Figref{fig:branes_O3_higgsing_C}, from which the $\orm(2)$ gauge theory is read. Note that this is $\orm(2)$ instead of $\sorm(2)$, as given in the rule in \eqref{rule:O_type_gauge}.}
\label{}
\end{figure}
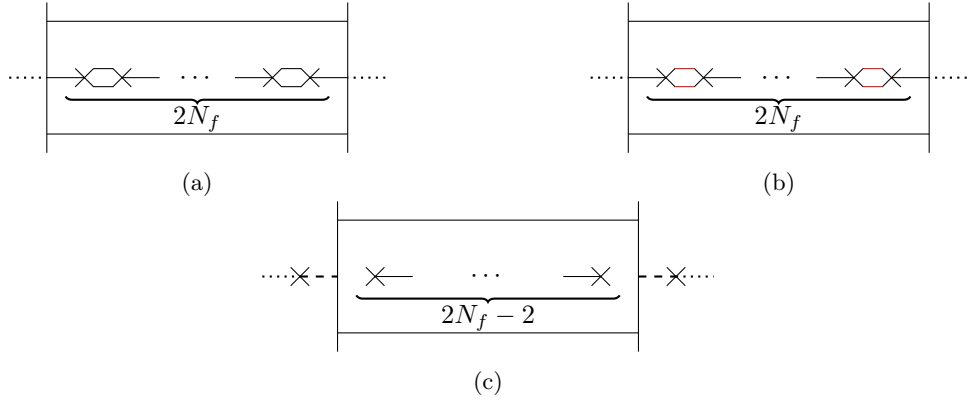

\begin{figure}[h!]
\centering
\begin{subfigure}{0.49\textwidth}
\centering
\begin{tikzpicture}
    \draw[dotted_brane] (-3,1)--(-2.5,1);
    \draw[dashed_brane] (-2.5,1)--(-2,1);
    \draw[dashed_brane] (2,1)--(2.5,1);
    \draw[dotted_brane] (2.5,1)--(3,1);
    \draw[-] (-2,0) -- (-2,2);
    \draw[-] (2,0) -- (2,2);
    \draw[-] (-2,1.75) -- (2,1.75);
    \draw[-] (-2,0.25) -- (2,0.25);
    \node (g1) at (-2.5,1) [D5] {};
    \node (g2) at (-1.5,1) [D5] {};
    \node (g3) at (2.5,1) [D5] {};
    \node (g4) at (1.5,1) [D5] {};
    \draw[-] (-1.5,1) -- (-1,1);
    \draw[-] (1,1) -- (1.5,1);
    \node at (0, 1) [scale=1]{$\cdots$} {};
    \draw [thick,decoration={brace,mirror,raise=0.25cm},decorate] (-1.75,1) -- (1.75,1);
    \node at (0,0.45) {$2N_f-2$};
\end{tikzpicture}
\caption{}
\label{fig:branes_O3_higgsing_ii_A}
\end{subfigure}
\begin{subfigure}{0.49\textwidth}
\centering
\begin{tikzpicture}
    \draw[dashed_brane] (-2.5,1)--(-2,1);
    \draw[dashed_brane] (2.5,1)--(2,1);
    \draw[dotted_brane] (-2.5,1)--(-3,1);
    \draw[dotted_brane] (2.5,1)--(3,1);
    \draw[-] (-2,0) -- (-2,2);
    \draw[-] (2,0) -- (2,2);
    \node (g1) at (-1.5,1) [D5] {};
    \node (g2) at (-1,1) [D5] {};
    \node (g3) at (1.5,1) [D5] {};
    \node (g4) at (1,1) [D5] {};
    \node (g5) at (-2.5,1) [D5] {};
    \node (g6) at (2.5,1) [D5] {};
    \draw[-] (-1,1) -- (-1.5,1);
    \draw[-] (1.5,1) -- (1,1);
    \node at (0, 1) [scale=1]{$\cdots$} {};
    \draw [thick,decoration={brace,mirror,raise=0.25cm},decorate] (-1.75,1) -- (1.75,1);
    \node at (0,0.45) {$2N_f-2$};
    \draw[-] (-1.62,1.12)--(-2,1.12);
    \draw[-] (-1.62,0.88)--(-2,0.88);
    \draw[-] (-0.88,1.12)--(-0.5,1.12);
    \draw[-] (-0.88,0.88)--(-0.5,0.88);
    \draw[-] (-1.62,1.12)--(-2,1.12);
    \draw[-] (-1.62,0.88)--(-2,0.88);
    \draw[-,color=red] (-1.38,1.12)--(-1.12,1.12);
    \draw[-,color=red] (-1.38,0.88)--(-1.12,0.88);
    \draw[-,color=red] (1.38,1.12)--(1.12,1.12);
    \draw[-,color=red] (1.38,0.88)--(1.12,0.88);
    \draw[-] (1.62,0.88)--(2,0.88);
    \draw[-] (1.62,1.12)--(2,1.12);
    \draw[-] (0.5,0.88)--(0.88,0.88);
    \draw[-] (0.5,1.12)--(0.88,1.12);
\end{tikzpicture}
\caption{}
\label{fig:branes_O3_higgsing_ii_B}
\end{subfigure}
\begin{subfigure}{0.49\textwidth}
\centering
\begin{tikzpicture}
    \draw[dotted_brane] (-2.5,1)--(-2,1);
    \draw[dotted_brane] (2.5,1)--(2,1);
    \draw[dotted_brane] (-3.5,1)--(-3,1);
    \draw[dotted_brane] (3.5,1)--(3,1);
    \draw[dashed_brane] (-2.5,1)--(-3,1);
    \draw[dashed_brane] (2.5,1)--(3,1);
    \draw[-] (-2,0) -- (-2,2);
    \draw[-] (2,0) -- (2,2);
    \node (g1) at (-1.5,1) [D5] {};
    \node (g2) at (-1,1) [D5] {};
    \node (g3) at (1.5,1) [D5] {};
    \node (g4) at (1,1) [D5] {};
    \node (g5) at (-3,1) [D5] {};
    \node (g6) at (-2.5,1) [D5] {};
    \node (g7) at (2.5,1) [D5] {};
    \node (g8) at (3,1) [D5] {};
    \draw[-] (-2,1) -- (-1.5,1);
    \draw[-] (-1,1) -- (-0.5,1);
    \draw[-] (0.5,1) -- (1,1);
    \draw[-] (1.5,1) -- (2,1);
    \node at (0, 1) [scale=1]{$\cdots$} {};
    \draw [thick,decoration={brace,mirror,raise=0.25cm},decorate] (-1.75,1) -- (1.75,1);
    \node at (0,0.45) {$2N_f-4$};
    \draw[-] (-1.38,1.12)--(-1.12,1.12);
    \draw[-] (-1.38,0.88)--(-1.12,0.88);
    \draw[-] (1.38,1.12)--(1.12,1.12);
    \draw[-] (1.38,0.88)--(1.12,0.88);
\end{tikzpicture}
\caption{}
\label{fig:branes_O3_higgsing_ii_C}
\end{subfigure}
\caption{The Higgsing process taking the $\orm(2)$ gauge theory with $N_f-1$ flavours, read from the brane system in \Figref{fig:branes_O3_higgsing_ii_A}, to the $\orm(1)$ gauge theory with $N_f-2$ flavours in \Figref{fig:branes_O3_higgsing_ii_C}. Splitting the $\frac{1}{2}$D3-brane along the fivebranes, shown in \Figref{fig:branes_O3_higgsing_ii_B}, allows for the segments coloured red to be `removed' under the Kraft-Procesi transition, which gives a brane interpretation of the lost $N_f-1$ moduli. The remaining segments either play the role of moduli or, in the case of the $\frac{1}{2}$D3-branes suspended between the $\frac{1}{2}$NS5-branes and $\frac{1}{2}$D5-branes, can be annihilated under a Hanany-Witten transition.}
\label{fig:branes_O3_higgsing_ii}
\end{figure}
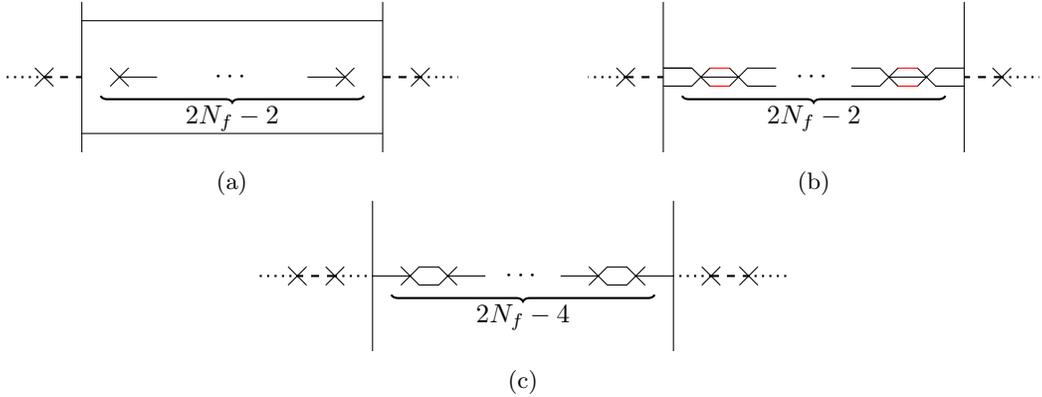

Importantly, the branes involved in the Krasft-Procesi transition, shown in \Figref{fig:branes_O3_higgsing_B}, locally support an $\orm(1)$ gauge theory with $N_f$ flavours, whose Higgs branch is the closure of the minimal nilpotent orbit of $\sprm(N_f)$. Further Higgsing of the theory can be realised in the brane system via the Kraft-Procesi transition shown in \Figref{fig:branes_O3_higgsing_ii}; splitting the $\frac{1}{2}$D5-branes and performing a Hanany-Witten transition results in the brane system in \Figref{fig:branes_O3_higgsing_ii_C}, which supports an $\orm(1)$ gauge theory with $N_f-2$ flavours. Clearly, this agrees with the field theory Hasse diagram given in \Figref{hasse:O3_Nf}.

Now consider the Coulomb branches of the $\orm(3)$ and $\orm(2)$ gauge theories given here. It is straightforward to show using the monopole formula \cite{Cremonesi:2013lqa, Hanany:2016pfm},
\begin{equation}
    \text{HS}(t) = \sum_{m\in \Gamma^{G^{\vee}}_{\mathcal{W}}}P_{G}(t; m)t^{2\Delta(t; m)},
\label{eq:monopole_formula}
\end{equation}
that both spaces have identical unrefined Hilbert series. Since the magnetic charges associated to $\orm(2k)$ and $\sorm(2k+1)$ take values on the same lattice (and the two groups have the same dressing factor $P_{G}$) it is sufficient to check that the conformal dimensions of the two quivers are the same. A simple calculation shows that in both cases $\Delta = (n-2)\left|m\right|$ where $m$ is the magnetic charge associated to (each) gauge group. As such, the unrefined Hilbert series of the Coulomb branches of the $\orm(3)$ and $\orm(2)$ gauge theories in \Figref{hasse:O3_Nf} coincide. Although this is not conclusive evidence that the two moduli spaces are in fact identical, it is significant evidence towards this conclusion. For this reason, results of this kind are used throughout this paper to motivate conjectures of Coulomb branch equivalence.

This example is included to provide a context for the manipulations performed in the rest of this paper. Using Kraft-Procesi transitions of the kind evidenced here between the $\orm(3)$ and $\orm(2)$ gauge theories, a set of identities on the Coulomb branches of various quivers can be found that, under certain conditions, swap $\sorm(2k+1)$ and $\orm(2k)$ gauge groups along with a compensating change to flavours. These transformations have the interpretation of a Higgsing that keeps the Coulomb branch invariant which, from the perspective of the Hasse diagram, motivates the conjecture that the theories related by the Higgsing form part of the same special piece \cite{juteau2023minimal,2023arXiv230807398F,Generic_singularities,achar_lcq,SOMMERS2001790}.
\section{The \boldmath$B_n \leftrightarrow \orm(2n)$ \unboldmath Identity}
\label{sec:Bn_O2n_Identity}
The observation in Section \ref{sec:motivating_example_O3_Nf} that the $\orm(3)$ and $\orm(2)$ gauge theories have the same Coulomb branch can be understood in terms of a broader set of Coulomb branch identities. These concern linear orthosymplectic quivers with a framed $B$-type node between two framed nodes of $C$-type. A similar argument to that given in Section \ref{sec:motivating_example_O3_Nf} shows the unrefined Hilbert series of the quivers' Coulomb branches to be identical, motivating the conjecture that the moduli spaces themselves are the same. As in Section \ref{sec:motivating_example_O3_Nf}, the Coulomb branch identities introduced in this section are conjectured to admit a brane interpretation as a Higgsing inside a special piece. Performing a Kraft-Procesi transition on the D3-branes created when a D5-brane splits along an $\widetilde{O3^{-}}$ plane makes the link between the two theories manifest, and provides further justification for the rule given in Section \ref{sec:Rule_OvsSO}.

Consider the theory given on the left in \eqref{quiv:identity:general}. The general type of manipulation considered in this section consists in changing the $\mathfrak{b}_{k}$ gauge node into the $\orm(2k)$ node shown on the right-hand side. Alongside a compensating change to the flavour groups $F_1$ and $F_2$, this action is conjectured to leave the Coulomb branch invariant. Current methods limit confidence in this conjecture to the unrefined Hilbert series, which is shown explicitly in Appendix \ref{app:proof_of_identities} to remain invariant. Although the intuition behind \eqref{quiv:identity:general} is made clearest using the brane interpretation given in Section \ref{sec:As_A_Higgsing}, the following considers a few generic examples.

In the examples below, the 
\begin{equation}
    \begin{tikzpicture}
      \node (a) at (0,0) {$\begin{tikzpicture}
        \node (1) [gauge, label=below:{$\mathfrak{b}_{k}$}] at (0,0) {};
        \node (2) [gauge, label=below:{$\mathfrak{c}_{k_3}$}] at (1,0) {};
        \node (3) [flavor, label=above:{$\orm(r_2)$}] at (1,1) {};
        \node (4) [flavor, label=above:{$\mathfrak{c}_{k_2}$}] at (0,1) {};
        \node (5) [gauge, label=below:{$\mathfrak{c}_{k_1}$}] at (-1,0) {};
        \node (6) [flavor, label=above:{$\orm(r_1)$}] at (-1,1) {};
        \draw (4)--(1)--(2)--(3)--(2)--(1)--(5)--(6);
        \draw[-] (-1.7,0) -- (5);
        \draw[-] (1.7,0) -- (2);
        \node (7) at (-2,-0.03) {$\cdots$};
        \node (8) at (2,-0.03) {$\cdots$};
        \end{tikzpicture}$};
    \node (b) at (4,0) {$\begin{tikzpicture}
    \node (1) at (0,0) [scale=1.25]{$\stackrel{\mathcal{C}}{=}$};
    \end{tikzpicture}$};
    \node (c) at (8,0) {$\begin{tikzpicture}
        \node (1) [gauge, label=below:{$\orm(2k)$}] at (0,0) {};
        \node (2) [gauge, label=below:{$\mathfrak{c}_{k_3}$}] at (1,0) {};
        \node (3) [flavor, label=right:{$\orm(r_2+1)$}] at (1,1) {};
        \node (4) [flavor, label=above:{$\mathfrak{c}_{k_2-1}$}] at (0,1) {};
        \node (5) [gauge, label=below:{$\mathfrak{c}_{k_1}$}] at (-1,0) {};
        \node (6) [flavor, label=left:{$\orm(r_1+1)$}] at (-1,1) {};
        \draw (4)--(1)--(2)--(3)--(2)--(1)--(5)--(6);
        \draw[-] (-1.7,0) -- (5);
        \draw[-] (1.7,0) -- (2);
        \node (7) at (-2,-0.03) {$\cdots$};
        \node (8) at (2,-0.03) {$\cdots$};
        \end{tikzpicture}$};
    \end{tikzpicture}
    \label{quiv:identity:general}
\end{equation}
\paragraph{Examples}
\begin{itemize}
    \item \boldmath$r_1=2m$, $r_2 = 2n+1$\unboldmath: One such example involves the two quivers in \eqref{quiv:identity:m=n_general}. Replacing the central $\mathfrak{b}_{k}$ gauge node with $\orm(2k)$ maintains the Coulomb banch as long as the flavours of the three gauge nodes are amended as shown. In the example in \eqref{quiv:identity:m=n_general}, the two $C$-type gauge nodes on either side see different flavour groups --- the identity remains valid if the two flavour groups are of the same type (in other words, the flavour groups of the two $C$-type gauge nodes could both be of type $D$ or type $B$ - the example in \eqref{quiv:identity:m=n_general} is included to illustrate how each of the $D$- and $B$-type flavours change).
\begin{equation}
    \begin{tikzpicture}
      \node (a) at (0,0) {$\begin{tikzpicture}
        \node (1) [gauge, label=below:{$\mathfrak{b}_{k}$}] at (0,0) {};
        \node (2) [gauge, label=below:{$\mathfrak{c}_{k_3}$}] at (1,0) {};
        \node (3) [flavor, label=above:{$\mathfrak{b}_n$}] at (1,1) {};
        \node (4) [flavor, label=above:{$\mathfrak{c}_{k_2}$}] at (0,1) {};
        \node (5) [gauge, label=below:{$\mathfrak{c}_{k_1}$}] at (-1,0) {};
        \node (6) [flavor, label=above:{$\mathfrak{d}_m$}] at (-1,1) {};
        \draw (4)--(1)--(2)--(3)--(2)--(1)--(5)--(6);
        \draw[-] (-1.7,0) -- (5);
        \draw[-] (1.7,0) -- (2);
        \node (7) at (-2,-0.03) {$\cdots$};
        \node (8) at (2,-0.03) {$\cdots$};
        \end{tikzpicture}$};
    \node (b) at (4,0) {$\begin{tikzpicture}
    \node (1) at (0,0) [scale=1.25]{$\stackrel{\mathcal{C}}{=}$};
    \end{tikzpicture}$};
    \node (c) at (8,0) {$\begin{tikzpicture}
        \node (1) [gauge, label=below:{$\orm(2k)$}] at (0,0) {};
        \node (2) [gauge, label=below:{$\mathfrak{c}_{k_3}$}] at (1,0) {};
        \node (3) [flavor, label=above:{$\mathfrak{d}_{n+1}$}] at (1,1) {};
        \node (4) [flavor, label=above:{$\mathfrak{c}_{k_2-1}$}] at (0,1) {};
        \node (5) [gauge, label=below:{$\mathfrak{c}_{k_1}$}] at (-1,0) {};
        \node (6) [flavor, label=above:{$\mathfrak{b}_{m}$}] at (-1,1) {};
        \draw (4)--(1)--(2)--(3)--(2)--(1)--(5)--(6);
        \draw[-] (-1.7,0) -- (5);
        \draw[-] (1.7,0) -- (2);
        \node (7) at (-2,-0.03) {$\cdots$};
        \node (8) at (2,-0.03) {$\cdots$};
        \end{tikzpicture}$};
    \end{tikzpicture}
    \label{quiv:identity:m=n_general}
\end{equation}
    \item \boldmath$r_1=r_2=1$\unboldmath: The $m=n=0$ case is an illustrative example in which the $\mathfrak{b}_{0}$ nodes on either side of the central $\sorm(2k+1)$ gauge node are promoted to $\mathfrak{d}_{1}$.
    \begin{equation}
    \begin{tikzpicture}
      \node (a) at (0,0) {$\begin{tikzpicture}
        \node (1) [gauge, label=below:{$\mathfrak{b}_{k}$}] at (0,0) {};
        \node (2) [gauge, label=below:{$\mathfrak{c}_{k_3}$}] at (1,0) {};
        \node (3) [flavor, label=above:{$\mathfrak{b}_0$}] at (1,1) {};
        \node (4) [flavor, label=above:{$\mathfrak{c}_{k_2}$}] at (0,1) {};
        \node (5) [gauge, label=below:{$\mathfrak{c}_{k_1}$}] at (-1,0) {};
        \node (6) [flavor, label=above:{$\mathfrak{b}_0$}] at (-1,1) {};
        \draw (4)--(1)--(2)--(3)--(2)--(1)--(5)--(6);
        \draw[-] (-1.7,0) -- (5);
        \draw[-] (1.7,0) -- (2);
        \node (7) at (-2,-0.03) {$\cdots$};
        \node (8) at (2,-0.03) {$\cdots$};
        \end{tikzpicture}$};
    \node (b) at (4,0) {$\begin{tikzpicture}
    \node (1) at (0,0) [scale=1.25]{$\stackrel{\mathcal{C}}{=}$};
    \end{tikzpicture}$};
    \node (c) at (8,0) {$\begin{tikzpicture}
        \node (1) [gauge, label=below:{$\orm(2k)$}] at (0,0) {};
        \node (2) [gauge, label=below:{$\mathfrak{c}_{k_3}$}] at (1,0) {};
        \node (3) [flavor, label=above:{$\mathfrak{d}_1$}] at (1,1) {};
        \node (4) [flavor, label=above:{$\mathfrak{c}_{k_2-1}$}] at (0,1) {};
        \node (5) [gauge, label=below:{$\mathfrak{c}_{k_1}$}] at (-1,0) {};
        \node (6) [flavor, label=above:{$\mathfrak{d}_1$}] at (-1,1) {};
        \draw (4)--(1)--(2)--(3)--(2)--(1)--(5)--(6);
        \draw[-] (-1.7,0) -- (5);
        \draw[-] (1.7,0) -- (2);
        \node (7) at (-2,-0.03) {$\cdots$};
        \node (8) at (2,-0.03) {$\cdots$};
        \end{tikzpicture}$};
    \end{tikzpicture}
    \label{quiv:identity:m=n=1}
\end{equation}
    \item \boldmath$r_1=r_2=0$ \unboldmath: This example illustrates the case in which both $C$-type nodes beside the central $B$-type node are initially unframed. Performing the transition causes them to pick up a $\mathfrak{b}_{0}$ framing --- the rank of the flavour group associated to the central $B$-type gauge node decreases by one.
    \begin{equation}
    \begin{tikzpicture}
      \node (a) at (0,0) {$\begin{tikzpicture}
        \node (1) [flavor, label=above:{$\mathfrak{c}_{k_2}$}] at (0,1) {};
        \node (2) [gauge, label=below:{$\mathfrak{b}_{k}$}] at (0,0) {};
        \node (3) [gauge, label=below:{$\mathfrak{c}_{k_3}$}] at (1,0) {};
        \node (4) [gauge, label=below:{$\mathfrak{c}_{k_1}$}] at (-1,0) {};
        \draw[-] (-1.7,0) -- (4);
        \draw[-] (1.7,0) -- (3);
        \node (5) at (-2,-0.03) {$\cdots$};
        \node (6) at (2,-0.03) {$\cdots$};
        \draw (1)--(2)--(3)--(2)--(4);
        \end{tikzpicture}$};
    \node (b) at (4,0) {$\begin{tikzpicture}
    \node (1) at (0,0) [scale=1.25]{$\stackrel{\mathcal{C}}{=}$};
    \end{tikzpicture}$};
    \node (c) at (8,0) {$\begin{tikzpicture}
        \node (1) [gauge, label=below:{$\orm(2k)$}] at (0,0) {};
        \node (2) [gauge, label=below:{$\mathfrak{c}_{k_3}$}] at (1,0) {};
        \node (3) [flavor, label=above:{$\mathfrak{b}_0$}] at (1,1) {};
        \node (4) [flavor, label=above:{$\mathfrak{c}_{k_2-1}$}] at (0,1) {};
        \node (5) [gauge, label=below:{$\mathfrak{c}_{k_1}$}] at (-1,0) {};
        \node (6) [flavor, label=above:{$\mathfrak{b}_0$}] at (-1,1) {};
        \draw[-] (-1.7,0) -- (5);
        \draw[-] (1.7,0) -- (2);
        \node (7) at (-2,-0.03) {$\cdots$};
        \node (8) at (2,-0.03) {$\cdots$};
        \draw (4)--(1)--(2)--(3)--(2)--(1)--(5)--(6);
        \end{tikzpicture}$};
    \end{tikzpicture}
    \label{quiv:identity:m=n=0}
    \end{equation}
\end{itemize}
As the above examples show, the general pattern for the flavours $F_1$ and $F_2$ in \eqref{quiv:identity:general} is $\sorm(r) \rightarrow \sorm(r+1)$. Alongside the diminution in rank of the $\mathfrak{c}_{k_2}$ flavour node, this hints at the possibility of interpreting the identity \eqref{quiv:identity:general} in terms of a process in a brane system, explored further in the next section.
\subsection{As a Higgsing}
\label{sec:As_A_Higgsing}
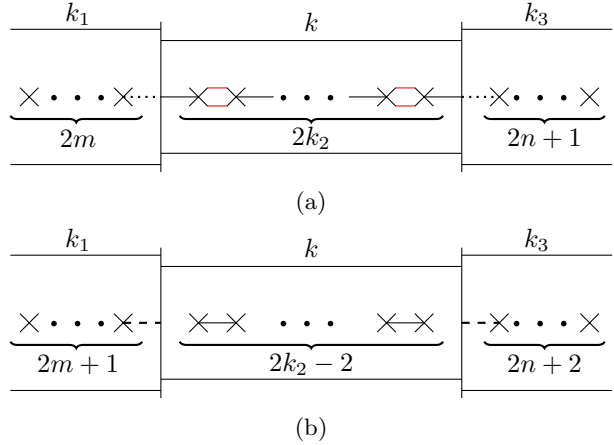
\begin{figure}[h!]
\centering
\begin{subfigure}{0.8\textwidth}
\centering
\begin{tikzpicture}
    \draw[-] (-4, 1.9) -- (-2,1.9);
    \node (f1) at (-2.5,1) [D5] {};
    \node (f2) at (-3.75,1) [D5] {};
    \node at (-3.05, 0.97) [scale=2]{$\cdots$} {};
    \draw [thick,decoration={brace,mirror,raise=0.25cm},decorate] (-4,1) -- (-2.25,1);
    \node at (-3.1,0.45) {$2m$};
    \draw[-] (-4, 0.1) -- (-2,0.1);
    \draw[dotted_brane] (-2.5,1) -- (-2,1);
    \draw[-] (-2,0) -- (-2,2);
    \draw[-] (2,0) -- (2,2);
    \node (h1) at (2.5,1) [D5] {};
    \node at (3.1, 0.97) [scale=2]{$\cdots$} {};
    \draw [thick,decoration={brace,mirror,raise=0.25cm},decorate] (2.3,1) -- (3.9,1);
    \node at (3.1,0.45) {$2n+1$};
    \node (h1) at (3.7,1) [D5] {};
    \draw[-] (2,1.9) -- (4,1.9);
    \draw[-] (2,0.1) -- (4,0.1);
    \draw[-] (-2,1.75) -- (2,1.75);
    \draw[-] (-2,0.25) -- (2,0.25);
    \node (g1) at (-1.5,1) [D5] {};
    \node (g2) at (-1,1) [D5] {};
    \node (g3) at (1.5,1) [D5] {};
    \node (g4) at (1,1) [D5] {};
    \draw[-,color=red] (-1.38,1.12)--(-1.12, 1.12);
    \draw[-,color=red] (-1.38,0.88)--(-1.12, 0.88);
    \draw[-,color=red] (1.38,1.12)--(1.12, 1.12);
    \draw[-,color=red] (1.38,0.88)--(1.12, 0.88);
    \draw[-] (-2,1) -- (-1.5,1);
    \draw[-] (-1,1) -- (-0.5,1);
    \draw[-] (0.5,1) -- (1,1);
    \draw[-] (1.5,1) -- (2,1);
    \draw[dotted_brane] (2,1) -- (2.5,1);
    \node at (0,2) {$k$};
    \node at (3,2.1) {$k_3$};
    \node at (-3.1,2.1) {$k_1$};
    \node at (0, 0.97) [scale=2]{$\cdots$} {};
    \draw [thick,decoration={brace,mirror,raise=0.25cm},decorate] (-1.75,1) -- (1.75,1);
    \node at (0,0.45) {$2k_2$};
\end{tikzpicture}
\caption{}
\label{fig:generic_identity_brane_interp_A}
\end{subfigure}
\begin{subfigure}{0.8\textwidth}
\centering
\begin{tikzpicture}
    \draw[-] (-4, 1.9) -- (-2,1.9);
    \node (f1) at (-2.5,1) [D5] {};
    \node (f2) at (-3.75,1) [D5] {};
    \node at (-3.05, 0.97) [scale=2]{$\cdots$} {};
    \draw [thick,decoration={brace,mirror,raise=0.25cm},decorate] (-4,1) -- (-2.25,1);
    \node at (-3.1,0.45) {$2m+1$};
    \draw[-] (-4, 0.1) -- (-2,0.1);
    \draw[dashed_brane] (-2.5,1) -- (-2,1);
    \draw[-] (-2,0) -- (-2,2);
    \draw[-] (2,0) -- (2,2);
    \node (h1) at (2.5,1) [D5] {};
    \node at (3.1, 0.97) [scale=2]{$\cdots$} {};
    \draw [thick,decoration={brace,mirror,raise=0.25cm},decorate] (2.3,1) -- (3.9,1);
    \node at (3.1,0.45) {$2n+2$};
    \node (h1) at (3.7,1) [D5] {};
    \draw[-] (2,1.9) -- (4,1.9);
    \draw[-] (2,0.1) -- (4,0.1);
    \draw[-] (-2,1.75) -- (2,1.75);
    \draw[-] (-2,0.25) -- (2,0.25);
    \node (g2) at (-1.5,1) [D5] {};
    \node (g4) at (1.5,1) [D5] {};
    \node (g5) at (-1,1) [D5] {};
    \node (g6) at (1,1) [D5] {};
    \draw[-] (-1.5,1) -- (-1,1);
    \draw[-] (1,1) -- (1.5,1);
    \draw[dashed_brane] (2,1) -- (2.5,1);
    \node at (0,2) {$k$};
    \node at (3,2.1) {$k_3$};
    \node at (-3.1,2.1) {$k_1$};
    \node at (0, 0.97) [scale=2]{$\cdots$} {};
    \draw [thick,decoration={brace,mirror,raise=0.25cm},decorate] (-1.75,1) -- (1.75,1);
    \node at (0,0.45) {$2k_2-2$};
\end{tikzpicture}
\caption{}
\label{fig:generic_identity_brane_interp_B}
\end{subfigure}
\caption{The Higgsing of the theory of \eqref{quiv:identity:m=n_general} to that on the right. The red branes in \Figref{fig:generic_identity_brane_interp_A} are removed via the Kraft-Procesi transition and the two D5-branes closest to the NS5-branes are HW-transitioned into the neighbouring interval to recover the configuration in \Figref{fig:generic_identity_brane_interp_B}. The transition is $c_{k_2}$ and is conjectured to be a `special Higgsing' (note that this terminology is not standard and is not desired for wider use) that remains in the same special piece in the Higgs branch Hasse diagram.}
\label{fig:generic_identity_brane_interp}
\end{figure}

Interpreting the conjectures of the previous section in terms of branes follows broadly the same logic as in Section \ref{sec:motivating_example_O3_Nf}. Like before, the two quivers' equivalent Coulomb branches are taken in the brane system to point towards a Higgsing between them that remains inside the same special piece in their Coulomb branch Hasse diagram. Consider for example the brane system given in \Figref{fig:generic_identity_brane_interp_A}, from which the quiver on the left in \eqref{quiv:identity:m=n_general} can be read. Removing the $\frac{1}{2}$-branes in red in \Figref{fig:generic_identity_brane_interp_A} under a Kraft-Procesi transition and performing an HW-transition on the left- and right-most $\frac{1}{2}$D5-branes results in the brane configuration in \Figref{fig:generic_identity_brane_interp_B}. Using the derivation rule of Section \ref{Section:Derivation_Rules}, the central $\frac{1}{2}$NS5-brane interval in \Figref{fig:generic_identity_brane_interp_B} appears to support an $\orm(2k)$ gauge theory, as on the right-hand side of \eqref{quiv:identity:m=n_general}.
That the Higgsing in \Figref{fig:generic_identity_brane_interp} preserves the Coulomb branch can be argued from the fact that no Coulomb branch moduli appear to be lost in the brane system. Note that this is only the case if the gauge node with $D$-type algebra is interpreted as $\orm(2k)$ instead of $\sorm(2k)$, in line with the proposal of Section \ref{sec:Rule_OvsSO}.

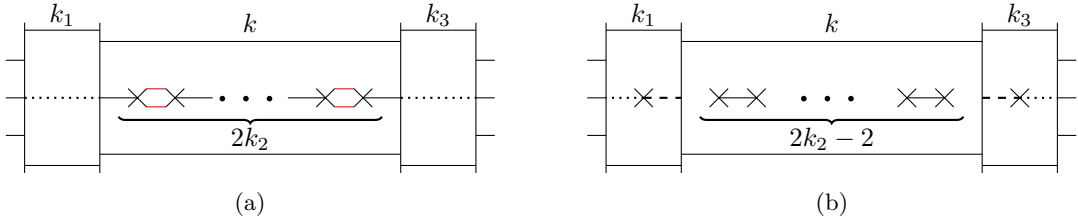
\begin{figure}[h!]
\centering
\begin{subfigure}{0.49\textwidth}
\centering
\begin{tikzpicture}
    \draw[-] (-3, 0) -- (-3,2);
    \draw[-] (-3, 1.5) -- (-3.25,1.5);
    \draw[-] (-3, 0.5) -- (-3.25,0.5);
    \draw[-] (3, 1.5) -- (3.25,1.5);
    \draw[-] (3, 0.5) -- (3.25,0.5);
    \draw[-] (3, 1) -- (3.25,1);
    \draw[-] (-3, 1) -- (-3.25,1);
    \draw[-] (3, 0) -- (3,2);
    \draw[-] (-3, 1.9) -- (-2,1.9);
    \draw[-] (-3, 0.1) -- (-2,0.1);
    \draw[dotted_brane] (-3,1) -- (-2,1);
    \draw[-] (-2,0) -- (-2,2);
    \draw[-] (2,0) -- (2,2);
    \draw[-] (2,1.9) -- (3,1.9);
    \draw[-] (2,0.1) -- (3,0.1);
    \draw[-] (-2,1.75) -- (2,1.75);
    \draw[-] (-2,0.25) -- (2,0.25);
    \node (g1) at (-1.5,1) [D5] {};
    \node (g2) at (-1,1) [D5] {};
    \node (g3) at (1.5,1) [D5] {};
    \node (g4) at (1,1) [D5] {};
    \draw[-,color=red] (-1.38,1.12)--(-1.12, 1.12);
    \draw[-,color=red] (-1.38,0.88)--(-1.12, 0.88);
    \draw[-,color=red] (1.38,1.12)--(1.12, 1.12);
    \draw[-,color=red] (1.38,0.88)--(1.12, 0.88);
    \draw[-] (-2,1) -- (-1.5,1);
    \draw[-] (-1,1) -- (-0.5,1);
    \draw[-] (0.5,1) -- (1,1);
    \draw[-] (1.5,1) -- (2,1);
    \draw[dotted_brane] (2,1) -- (3,1);
    \node at (0,2) {$k$};
    \node at (2.5,2.1) {$k_3$};
    \node at (-2.5,2.1) {$k_1$};
    \node at (0, 0.97) [scale=2]{$\cdots$} {};
    \draw [thick,decoration={brace,mirror,raise=0.25cm},decorate] (-1.75,1) -- (1.75,1);
    \node at (0,0.45) {$2k_2$};
\end{tikzpicture}
\caption{}
\label{fig:m=n=0_identity_brane_interp_A}
\end{subfigure}
\begin{subfigure}{0.49\textwidth}
\centering
\begin{tikzpicture}
   \draw[-] (-3, 0) -- (-3,2);
    \draw[-] (-3, 1.5) -- (-3.25,1.5);
    \draw[-] (-3, 0.5) -- (-3.25,0.5);
    \draw[-] (3, 1.5) -- (3.25,1.5);
    \draw[-] (3, 0.5) -- (3.25,0.5);
    \draw[-] (3, 1) -- (3.25,1);
    \draw[-] (-3, 1) -- (-3.25,1);
    \draw[-] (3, 0) -- (3,2);
    \draw[-] (-3, 1.9) -- (-2,1.9);
    \draw[-] (-3, 0.1) -- (-2,0.1);
    \draw[dashed_brane] (-2.5,1) -- (-2,1);
    \draw[dotted_brane] (-2.5,1) -- (-3,1);
    \draw[-] (-2,0) -- (-2,2);
    \draw[-] (2,0) -- (2,2);
    \draw[-] (2,1.9) -- (3,1.9);
    \draw[-] (2,0.1) -- (3,0.1);
    \draw[-] (-2,1.75) -- (2,1.75);
    \draw[-] (-2,0.25) -- (2,0.25);
    \node (g1) at (-2.5,1) [D5] {};
    \node (g2) at (-1,1) [D5] {};
    \node (g2) at (-1.5,1) [D5] {};
    \node (g3) at (2.5,1) [D5] {};
    \node (g4) at (1,1) [D5] {};
    \node (g4) at (1.5,1) [D5] {};
    \draw[-] (-1,1) -- (-1.5,1);
    \draw[-] (1.5,1) -- (1,1);
    \draw[dashed_brane] (2,1) -- (2.5,1);
    \draw[dotted_brane] (2.5,1) -- (3,1);
    \node at (0,2) {$k$};
    \node at (2.5,2.1) {$k_3$};
    \node at (-2.5,2.1) {$k_1$};
    \node at (0, 0.97) [scale=2]{$\cdots$} {};
    \draw [thick,decoration={brace,mirror,raise=0.25cm},decorate] (-1.75,1) -- (1.75,1);
    \node at (0,0.45) {$2k_2-2$};
\end{tikzpicture}
\caption{}
\label{fig:m=n=0_identity_brane_interp_B}
\end{subfigure}
\caption{The brane interpretation of the identity in \eqref{quiv:identity:m=n=0}. In \Figref{fig:m=n=0_identity_brane_interp_A}, which corresponds to the theory on the left of \eqref{quiv:identity:m=n=0}, the red branes are removed under a Kraft-Procesi transition. Rearranging the brane system using an HW-transition results in that given in \Figref{fig:m=n=0_identity_brane_interp_B}. Using the rule in Section \ref{sec:Rule_OvsSO}, the central NS5-brane interval is interpreted as supporting an $\orm(2k)$ gauge theory. The placements of the $\frac{1}{2}$D5-branes exactly correspond to the flavours on the right hand side of \eqref{quiv:identity:m=n=0}.}
\label{fig:m=n=0_identity_brane_interp}
\end{figure}

The phenomenon is essentially the same in the case of \eqref{quiv:identity:m=n=1} and \eqref{quiv:identity:m=n=0}. For \eqref{quiv:identity:m=n=0}, the brane realisation is given in \Figref{fig:m=n=0_identity_brane_interp}. The Kraft-Procesi transition deposits a single $\frac{1}{2}$D5-brane into each of the left- and right-hand side $\frac{1}{2}$NS5-brane intervals, as reflected in the $\mathfrak{b}_{0}$ flavours.
\subsection{Special Piece Generalisation}
As remarked in Section \ref{sec:Rule_OvsSO} and \cite{Cabrera:2017ucb}, in quivers whose Coulomb branch corresponds to a nilpotent orbit closure in the nilcone of a $B/C/D$-type algebra, the appearance of $\orm(2n)$ instead of $\sorm(2n)$ in gauge nodes can be identified with factors of $\mathbb{Z}_{2}$ in the Lusztig canonical quotient associated to the leaf.
\section{The \boldmath$BC$- and $CB$-Chains\unboldmath}
\label{sec:BC_Chains}
\begin{equation}
    \begin{tikzpicture}
      \node (a) at (0,0) {$\begin{tikzpicture}
        \node (1) [gauge, label=below:{$\mathfrak{b}_k$}] at (0,0) {};
        \node (2) [gauge, label=below:{$\mathfrak{c}_k$}] at (1,0) {};
        \node (3) [gauge, label=below:{$\mathfrak{c}_k$}] at (4,0) {};
        \node (4) [gauge, label=below:{$\mathfrak{b}_k$}] at (5,0) {};
        \node (5) [flavor, label=above:{$\mathfrak{c}_k$}] at (0,1) {};
        \node (6) [flavor, label=above:{$\mathfrak{c}_k$}] at (5,1) {};
        \draw (5) -- (1) -- (2);
        \draw (3) -- (4) -- (6);
        \draw (2) -- (2,0);
        \draw (3) -- (3,0);
        \node at (2.54, -0.03) [scale=2]{$\cdots$} {};
        \end{tikzpicture}$};
    \node (b) at (4,0) {$\begin{tikzpicture}
    \node (1) at (0,0) [scale=1.25]{$\stackrel{\mathcal{C}}{=}$};
    \end{tikzpicture}$};
    \node (c) at (8,0) {$\begin{tikzpicture}
        \node (1) [gauge, label=below:{$\mathfrak{c}_k$}] at (0,0) {};
        \node (2) [gauge, label=below:{$\mathfrak{b}_k$}] at (1,0) {};
        \node (3) [gauge, label=below:{$\mathfrak{b}_k$}] at (4,0) {};
        \node (4) [gauge, label=below:{$\mathfrak{c}_k$}] at (5,0) {};
        \node (5) [flavor, label=above:{$\mathfrak{b}_k$}] at (0,1) {};
        \node (6) [flavor, label=above:{$\mathfrak{b}_k$}] at (5,1) {};
        \draw (5) -- (1) -- (2);
        \draw (3) -- (4) -- (6);
        \draw (2) -- (2,0);
        \draw (3) -- (3,0);
        \node at (2.54, -0.03) [scale=2]{$\cdots$} {};
        \end{tikzpicture}$};
    \end{tikzpicture}
    \label{quiv:BC_Chain}
\end{equation}
Another identity introduced in this note concerns the $BC$- and $CB$-chain quivers in \eqref{quiv:BC_Chain}. The Coulomb branch, which is shared by both quivers, is of dimension $k(2N+1)$ and has global symmetry $\sorm(2N+2)$ -- Coulomb branch Hilbert series for various examples are given in Appendix \ref{app:BC_Chain_HS}.

Like the identities proposed in Section \ref{sec:Bn_O2n_Identity}, the equivalence of the two Coulomb branches in \eqref{quiv:BC_Chain} can be demonstrated at the level of unrefined Hilbert series using the monopole formula \cite{Cremonesi:2013lqa}, which calculates the Hilbert series for a moduli space of dressed monopole operators using a quiver's conformal dimension $\Delta(t;m)$, gauge group-dependent dressing factors $P_{G}(t;m)$ and magnetic lattice Weyl orbits $\Gamma^{G^{\vee}}_{\mathcal{W}}$. Since magnetic lattices and dressing factors are identical for both $B$- and $C$-type groups, to check the conjecture \eqref{quiv:BC_Chain} at the level of unrefined Hilbert series it suffices to match the two conformal dimensions.
In \Figref{fig:BC_Annotated}, \eqref{quiv:BC_Chain} is rewritten alongside the magnetic charges associated to each gauge group. First consider the $CB$-chain in \Figref{fig:BC_Annotated1} with conformal dimension $\Delta_{\text{L}}$ given in \eqref{eq:DeltaL_CB_chain}.

\begin{figure}[h!]
\centering
\begin{subfigure}{0.49\textwidth}
    \scalebox{.85}{\begin{tikzpicture}
        \node (1) [gauge, label=below:{$\mathfrak{c}_k$}] at (0,0) {};
        \node (2) [gauge, label=below:{$\mathfrak{b}_k$}] at (1,0) {};
        \node (3) [gauge, label=below:{$\mathfrak{b}_k$}] at (4,0) {};
        \node (4) [gauge, label=below:{$\mathfrak{c}_k$}] at (5,0) {};
        \node (5) [flavor, label=above:{$\mathfrak{b}_k$}] at (0,1) {};
        \node (6) [flavor, label=above:{$\mathfrak{b}_k$}] at (5,1) {};
        \draw (5) -- (1) -- (2);
        \draw (3) -- (4) -- (6);
        \draw (2) -- (2,0);
        \draw (3) -- (3,0);
        \node at (2.54, -0.03) [scale=2]{$\cdots$} {};
        \draw [thick,decoration={brace,mirror,raise=1.5cm},decorate] (-0.25,0) -- (5.25,0);
        \node at (2.5,-2) {$2N+1$};
        \node at (0,-1) {$m_{i}^{1}$};
        \node at (1,-1) {$n_{i}^{1}$};
        \node at (4,-1) {$n_{i}^{N}$};
        \node at (5,-1) {$m_{i}^{N+1}$};
        \node at (-2,-1) {$i=1,\cdots,k$};
        \end{tikzpicture}}
        \caption{}
        \label{fig:BC_Annotated1}
    \end{subfigure}
    \begin{subfigure}{0.49\textwidth}
    \scalebox{.85}{\begin{tikzpicture}
        \node (1) [gauge, label=below:{$\mathfrak{b}_k$}] at (0,0) {};
        \node (2) [gauge, label=below:{$\mathfrak{c}_k$}] at (1,0) {};
        \node (3) [gauge, label=below:{$\mathfrak{c}_k$}] at (4,0) {};
        \node (4) [gauge, label=below:{$\mathfrak{b}_k$}] at (5,0) {};
        \node (5) [flavor, label=above:{$\mathfrak{c}_k$}] at (0,1) {};
        \node (6) [flavor, label=above:{$\mathfrak{c}_k$}] at (5,1) {};
        \draw (5) -- (1) -- (2);
        \draw (3) -- (4) -- (6);
        \draw (2) -- (2,0);
        \draw (3) -- (3,0);
        \node at (2.54, -0.03) [scale=2]{$\cdots$} {};
        \draw [thick,decoration={brace,mirror,raise=1.5cm},decorate] (-0.25,0) -- (5.25,0);
        \node at (2.5,-2) {$2N+1$};
        \node at (0,-1) {$g_{i}^{1}$};
        \node at (1,-1) {$h_{i}^{1}$};
        \node at (4,-1) {$h_{i}^{N}$};
        \node at (5,-1) {$g_{i}^{N+1}$};
        \node at (7,-1) {$i=1,\cdots,k$};
        \end{tikzpicture}}
        \caption{}
        \label{fig:BC_Annotated2}
    \end{subfigure}
\caption{\Figref{fig:BC_Annotated1} gives the $CB$-quiver with $N+1$ gauge nodes of $C$-type and $N$ gauge nodes of $B$-type and \Figref{fig:BC_Annotated2} gives the $BC$-quiver with $N+1$ gauge nodes of $B$-type and $N$ gauge nodes of $C$-type. The tuple of magnetic charges associated to each gauge node is given below the algebra label --- in \Figref{fig:BC_Annotated1} $C$- and $B$-type nodes have charges of the form $m^{\alpha}_{i}$ and $n^{\alpha}_{i}$ respectively while in \Figref{fig:BC_Annotated2} they are labelled by $h^{\alpha}_{i}$ and $g^{\alpha}_{i}$.}
\label{fig:BC_Annotated}
\end{figure}

\begin{align}
    \Delta_{\text{L}}^{\text{Hyp}} = & k\sum_{i=1}^{k} \left| m_{i}^{1}\right|+k\sum_{i=1}^{k}\left|m_{i}^{N+1}\right| + \frac{1}{2}\sum_{\alpha=1}^{N}\sum_{i,j=1}^{k}\left( \left| m_{i}^{\alpha}+n_{j}^{\alpha}\right| + \left| m_{i}^{\alpha}-n_{j}^{\alpha}\right| \right)+\sum_{\alpha=1}^{N+1}\sum_{i=1}^{k}\left|m_{i}^{\alpha}\right|\\& +\frac{1}{2}\sum_{\alpha=1}^{N}\sum_{i,j=1}^{k}\left( \left| n_{i}^{\alpha}+m_{j}^{\alpha+1}\right| + \left| n_{i}^{\alpha}-m_{j}^{\alpha+1}\right| \right)\\
    \Delta_{\text{L}}^{\text{Vec}} = & -\sum_{\alpha=1}^{N+1}\sum_{i<j}^{k}\left(\left|m_{i}^{\alpha}+m_{j}^{\alpha}\right|+\left|m_{i}^{\alpha}-m_{j}^{\alpha}\right|\right)-\sum_{\alpha=1}^{N}\sum_{i<j}^{k}\left(\left|n_{i}^{\alpha}+n_{j}^{\alpha}\right|+\left|n_{i}^{\alpha}-n_{j}^{\alpha}\right|\right)\\& -2\sum_{\alpha=1}^{N+1}\sum_{i=1}^{k}\left|m_{i}^{\alpha}\right| -\sum_{\alpha=1}^{N}\sum_{i=1}^{k}\left|n_{i}^{\alpha}\right|\\
    \Delta_{\text{L}} =& \; \Delta_{\text{L}}^{\text{Hyp}}+\Delta_{\text{L}}^{\text{Vec}} = \; \Sigma_{m,n} -\sum_{\alpha=2}^{N}\sum_{i=1}^{k}\left|m_{i}^{\alpha}\right|-\sum_{\alpha=1}^{N}\sum_{i=1}^{k}\left|n_{i}^{\alpha}\right|+(k-1)\sum_{i=1}^{k} \left( \left| m_{i}^{1}\right|+\left|m_{i}^{N+1}\right| \right)
\label{eq:DeltaL_CB_chain}
\end{align}
Note that $\Sigma_{m,n}$ is a repackaging of terms of the form $\left|m\pm n\right|$, $\left|m_{i}\pm m_{j}\right|$ and $\left|n_{i}\pm n_{j}\right|$. The $BC$-chain in \Figref{fig:BC_Annotated2} with conformal dimension $\Delta_{\text{R}}$ proceeds analogously \eqref{eq:DeltaR_BC_chain}.
\begin{align}
    \Delta_{\text{R}}^{\text{Hyp}} = & k\sum_{i=1}^{k}\left|g^{1}_{i}\right|+k\sum_{i=1}^{k}\left|g_{i}^{N+1}\right|+\sum_{\alpha=1}^{N}\sum_{i=1}^{k}\left|h_{i}^{\alpha}\right| + \frac{1}{2}\sum_{\alpha=1}^{N}\sum_{i,j=1}^{k}\left( \left| g_{i}^{\alpha}+h_{j}^{\alpha}\right| + \left| g_{i}^{\alpha}-h_{j}^{\alpha}\right| \right)\\& \frac{1}{2}\sum_{\alpha=1}^{N}\sum_{i,j=1}^{k}\left( \left| h_{i}^{\alpha}+g_{j}^{\alpha+1}\right| + \left| h_{i}^{\alpha}-g_{j}^{\alpha+1}\right| \right)\\
    \Delta_{\text{R}}^{\text{Vec}} = & -\sum_{\alpha=1}^{N+1}\sum_{i<j}^{k}\left(\left|g_{i}^{\alpha}+g_{j}^{\alpha}\right|+\left|g_{i}^{\alpha}-g_{j}^{\alpha}\right|\right)-\sum_{\alpha=1}^{N}\sum_{i<j}^{k}\left(\left|h_{i}^{\alpha}+h_{j}^{\alpha}\right|+\left|h_{i}^{\alpha}-h_{j}^{\alpha}\right|\right)\\& -2\sum_{\alpha=1}^{N}\sum_{i=1}^{k}\left|h_{i}^{\alpha}\right| -\sum_{\alpha=1}^{N+1}\sum_{i=1}^{k}\left|g_{i}^{\alpha}\right|\\
    \Delta_{\text{R}} =& \Delta_{\text{R}}^{\text{Hyp}}+\Delta_{\text{R}}^{\text{Vec}} = \; \Sigma_{g,h} -\sum_{\alpha=2}^{N}\sum_{i=1}^{k}\left|g_{i}^{\alpha}\right|-\sum_{\alpha=1}^{N}\sum_{i=1}^{k}\left|h_{i}^{\alpha}\right|+(k-1)\sum_{i=1}^{k} \left( \left| g_{i}^{1}\right|+\left|g_{i}^{N+1}\right| \right)
\label{eq:DeltaR_BC_chain}
\end{align}
Where again $\Sigma_{g,h}$ packages terms of the form $\left|g\pm h\right|$, $\left|g_{i}\pm g_{j}\right|$ and $\left|h_{i}\pm h_{j}\right|$. Since $(m^{\alpha},n^{\beta})$ and $(g^{\alpha},h^{\beta})$, for $\alpha = 1,\cdots,N+1$, $\beta = 1,\cdots,N$, are summed over the same lattice, the conformal dimension contributions of the two quivers in \Figref{fig:BC_Annotated1} and \Figref{fig:BC_Annotated2} are identical.

Note that the identities in Section \ref{sec:Bn_O2n_Identity} are also applicable to the $BC$-quiver given in \eqref{quiv:BC_Chain}. One such example is given in \eqref{quiv:Both_Identities_Example}.
\begin{equation}
    \begin{tikzpicture}
      \node (a) at (0,0) {$\begin{tikzpicture}
        \node (1) [gauge, label=below:{$\mathfrak{b}_k$}] at (0,0) {};
        \node (2) [gauge, label=below:{$\mathfrak{c}_k$}] at (1,0) {};
        \node (3) [gauge, label=below:{$\mathfrak{c}_k$}] at (4,0) {};
        \node (4) [gauge, label=below:{$\mathfrak{b}_k$}] at (5,0) {};
        \node (5) [flavor, label=above:{$\mathfrak{c}_k$}] at (0,1) {};
        \node (6) [flavor, label=above:{$\mathfrak{c}_k$}] at (5,1) {};
        \draw (5) -- (1) -- (2);
        \draw (3) -- (4) -- (6);
        \draw (2) -- (2,0);
        \draw (3) -- (3,0);
        \node at (2.54, -0.03) [scale=2]{$\cdots$} {};
        \end{tikzpicture}$};
    \node (b) at (4,0) {$\begin{tikzpicture}
    \node (1) at (0,0) [scale=1.25]{$\stackrel{\mathcal{C}}{=}$};
    \end{tikzpicture}$};
    \node (c) at (8,0) {$\begin{tikzpicture}
        \node (1) [gauge, label=below:{$\orm(2k)$}] at (0,0) {};
        \node (2) [flavor, label=above:{$\mathfrak{c}_{k-1}$}] at (0,1) {};
        \node (3) [gauge, label=below:{$\mathfrak{c}_k$}] at (4,0) {};
        \node (4) [gauge, label=below:{$\orm(2k)$}] at (5,0) {};
        \node (5) [gauge, label=below:{$\mathfrak{c}_k$}] at (1,0) {};
        \node (6) [flavor, label=above:{$\mathfrak{c}_{k-1}$}] at (5,1) {};
        \node (7) [flavor, label=above:{$\mathfrak{b}_0$}] at (1,1) {};
        \node (8) [flavor, label=above:{$\mathfrak{b}_0$}] at (4,1) {};
        \draw[-] (1) -- (2) -- (1) -- (5) -- (7) -- (5) -- (2,0);
        \draw (3) -- (4) -- (6);
        \draw (3) -- (3,0);
        \draw[-] (8) -- (3);
        \node at (2.54, -0.03) [scale=2]{$\cdots$} {};
        \end{tikzpicture}$};
    \end{tikzpicture}
    \label{quiv:Both_Identities_Example}
\end{equation}
\section{Gauging \boldmath$\mathbb{Z}_{2}$ Flavour Subgroups of $\sprm(k)$ SQCD\unboldmath}
\label{sec:gauging_elec_magnet_perspec}
In three dimensions, the Higgs branch of SQCD with gauge group $\sprm(k)$ exhibits a range of different structures as $N_f$, the number of flavours, is varied. For instance, $N_f = 2k$ cleaves the Higgs branch into a union of two cones \cite{Ferlito:2016grh, Bourget:2023cgs} with global symmetry $\sorm(4k)$, while the Higgsing pattern for $N_f=2k+1$ is a series of $d_{2n+1}$ transitions. In this section, a set of Higgs and Coulomb branch relations will be presented for theories under the gauging of $\mathbb{Z}_{2}$ subgroups of their flavour symmetries.
\begin{equation}
\begin{tikzpicture}
    \node (1) [flavor, label=above:{$\mathfrak{d}_{N_f}$}] at (0,1) {};
    \node (2) [gauge, label=below:{$\mathfrak{c}_{k}$}] at (0,0) {};
    \draw (1)--(2);
\end{tikzpicture}
\label{SPk_SQCD}
\end{equation}
Consider the $\sprm(k)$ SQCD with $N_f=2k+1$ in \eqref{SPk_SQCD}. Gauging successive $\mathbb{Z}_{2}$ subgroups (denoted $\mathfrak{b}_{0}$) of the flavour symmetry that act trivially on monopole operators leaves the Coulomb branch unchanged, as shown in \eqref{quiv:gaugingZ2s}. \footnote{It is possible to stipulate that monopole operators are charged under this $\mathbb{Z}_{2}$, considered in \cite{Grimminger:2024mks}. In this case the Coulomb branch will generically change as new gauge-invariant combinations of monopole operators emerge. The authors thank Noppadol Mekareeya and William Harding for discussion on this point.} The embedding is specified as,
\begin{equation}
    [1,0,\cdots,0]_{D} \mapsto [1,0,\cdots,0]_{B} +[0]_{B},
\end{equation}
where the first term on the right-hand side transforms trivially under $\mathbb{Z}_{2}$ and the second non-trivially.
Such gaugings will generically change the Higgs branch. Using S-duality in a Type IIB brane system, this $\mathbb{Z}_{2}$ gauging is conjectured to be related under 3d mirror symmetry \footnote{Although Hilbert series checks suggest that the moduli spaces of the electric and magnetic theories in Table \ref{table:SQCD_with_gaugings_elec_mag} obey a 3d mirror symmetry relationship, this alone is insufficient to conclude that the theories are strictly dual.} to gauging the diagonal $\mathbb{Z}_{2}$ lattice symmetry on a maximal $DC$-chain, as shown in Table \ref{table:SQCD_with_gaugings_elec_mag}. This conjecture is supported by matching Coulomb branch Hilbert series of the magnetic theories of Table \ref{table:SQCD_with_gaugings_elec_mag} with those of the Higgs branches of the electric theories. The map between $\mathbb{Z}_{2}$ flavour gauging on the left-hand column and the diagonal $\mathbb{Z}_{2}$ on the right-hand column of Table \ref{table:SQCD_with_gaugings_elec_mag} is in some sense reminiscent of \cite{Kapustin_1999}. Note that information on this sort of diagonal gauging can be found in \cite{Cabrera:2017ucb}.
\begin{equation}
    \begin{tikzpicture}
    \node (a) at (0,0) {$\begin{tikzpicture}
        \node (1) [flavor, label=above:{$\mathfrak{d}_{2k+1}$}] at (0,1) {};
        \node (2) [gauge, label=below:{$\mathfrak{c}_{k}$}] at (0,0) {};
        \draw (1)--(2);
        \end{tikzpicture}$};
    \node (b) at (2,0) {$\begin{tikzpicture}
        \node (1) at (0,0) [scale=1.25]{$\stackrel{\mathcal{C}}{=}$};
        \end{tikzpicture}$};
    \node (c) at (5,0) {$\begin{tikzpicture}
        \node (1) [flavor, label=above:{$\mathfrak{d}_{2k+1-m}$}] at (0,1) {};
        \node (2) [gauge, label=below:{$\mathfrak{c}_{k}$}] at (0,0) {};
        \node (3) [gauge, label=right:{$\mathfrak{b}_{0}$}] at (1,0.75) {};
        \node (4) [gauge, label=right:{$\mathfrak{b}_{0}$}] at (1,-0.75) {};
        \node (5) at (1,0.2) [scale=1.25]{$\vdots$};
        \node (6) at (2.5,0) {$2m$};
        \draw [thick,decoration={brace},decorate, rotate=90] (1,-2) -- (-1,-2);
        \draw (1)--(2)--(3)--(2)--(4);
        \end{tikzpicture}$};
    \node (b) at (8,0) {$\begin{tikzpicture}
        \node (1) at (0,0) [scale=1.25]{$\stackrel{\mathcal{C}}{=}$};
        \end{tikzpicture}$};
    \node (c) at (11,0) {$\begin{tikzpicture}
        \node (1) [flavor, label=above:{$\mathfrak{b}_{2k-m}$}] at (0,1) {};
        \node (2) [gauge, label=below:{$\mathfrak{c}_{k}$}] at (0,0) {};
        \node (3) [gauge, label=right:{$\mathfrak{b}_{0}$}] at (1,0.75) {};
        \node (4) [gauge, label=right:{$\mathfrak{b}_{0}$}] at (1,-0.75) {};
        \node (5) at (1,0.2) [scale=1.25]{$\vdots$};
        \node (6) at (2.75,0) {$2m+1$};
        \draw [thick,decoration={brace},decorate, rotate=90] (1,-2) -- (-1,-2);
        \draw (1)--(2)--(3)--(2)--(4);
        \end{tikzpicture}$};
    \end{tikzpicture}
    \label{quiv:gaugingZ2s}
\end{equation}
Like in Section \ref{sec:As_A_Higgsing}, it is instructive to consider a brane interpretation of the diagonal gauging shown in Table \ref{table:SQCD_with_gaugings_elec_mag}. Take the initial brane configuration in \Figref{fig:SQCD_One_Z2_Gauging_A} corresponding to the electric theory in \Figref{fig:SQCD_One_Z2_Gauging_B}. Using a Type IIB brane system, the brane configuration can be related to the magnetic theory given in \Figref{fig:one_Z2_gauging_dual}, where now the $\frac{1}{2}$NS5-brane on the left-hand side of \Figref{fig:SQCD_One_Z2_Gauging_A} has become a $\frac{1}{2}$D5-brane. Using an analogue of the conjecture in Section \ref{sec:Rule_OvsSO}, it appears that $\frac{1}{2}$D5-branes book-ending a maximal $DC$-chain have the effect of diagonally gauging a $\mathbb{Z}_{2}$ lattice subgroup in the D3-brane worldvolume theory, as shown in \eqref{rule:chain_rule}.
\begin{equation}
        \centering
        \begin{tikzpicture}
            \node (a) at (0,0) {$\begin{tikzpicture}
                \draw[-] (-1,0) -- (-1,2);
                \draw[-] (0,0) -- (0,2);
                \node (g1) at (-1.5,1) [D5] {};
                \draw[-] (-1,1.5) -- (0,1.5);
                \draw[-] (-1,0.5) -- (0,0.5);
                \draw[-] (0,0.25) -- (1,0.25);
                \draw[-] (0,1.75) -- (1,1.75);
                \draw [dashed_brane] (0,1) -- (1,1);
                \draw [dashed_brane] (-1,1) -- (-1.5,1);
                \draw [dotted_brane] (-1.5,1) -- (-1.75,1);
                \node at (-0.5,1.8) {$1$};
                \node at (0.5,2) {$2$};
                \node at (1.5,1.8) {$3$};
                \draw[-] (1,0) -- (1,2);
                \draw[-] (2,0) -- (2,2);
                \draw[-] (1,1.5) -- (2,1.5);
                \draw[-] (1,0.5) -- (2,0.5);
                \draw[-] (2,1.75) -- (2.25,1.75);
                \draw[-] (2,0.25) -- (2.25,0.25);

                \draw[-] (6,0) -- (6,2);
                \draw[-] (7,0) -- (7,2);
                \node (g1) at (7.5,1) [D5] {};
                \draw[-] (7,0.25) -- (8,0.25);
                \draw[-] (7,1.75) -- (8,1.75);
                \draw[-] (6,1.5) -- (7,1.5);
                \draw[-] (6,0.5) -- (7,0.5);
                \draw[-] (5,0.25) -- (6,0.25);
                \draw[-] (5,1.75) -- (6,1.75);
                \draw [dashed_brane] (5,1) -- (6,1);
                \draw [dashed_brane] (7,1) -- (7.5,1);
                \draw [dotted_brane] (7.5,1) -- (7.75,1);
                \node at (6.5,1.8) {$n$};
                \node at (5.5,2) {$n-1$};
                \node at (4.5,1.8) {$n-1$};
                \node at (7.5,2) {$n$};
                \draw[-] (5,0) -- (5,2);
                \draw[-] (4,0) -- (4,2);
                \draw[-] (4,1.5) -- (5,1.5);
                \draw[-] (4,0.5) -- (5,0.5);
                \draw[-] (4,1.75) -- (3.75,1.75);
                \draw[-] (4,0.25) -- (3.75,0.25);
                \node at (3,1) [scale=2]{$\cdots$};
            \end{tikzpicture}$};
        \node (b) at (0,-3) {$\begin{tikzpicture}
            \node (1) [gauge, label=below:{$[\mathfrak{d}_1$}] at (0,0) {};
            \node (2) [gauge, label=below:{$\mathfrak{c}_{1}$}] at (1,0) {};
            \node (3) [gauge, label=below:{$\mathfrak{d}_{2}$}] at (2,0) {};
            \node (4) [gauge, label=below:{$\mathfrak{c}_{2}$}] at (3,0) {};
            \node (5) at (4, 0) [scale=1]{$\cdots$} {};
            \node (6) [gauge, label=below:{$\mathfrak{d}_{n}]$}] at (5,0) {};
            \node (7) [gauge, label=below:{$\mathfrak{c}_{n}$}] at (6,0) {};
            \node (8) [flavor, label=above:{$F$}] at (6,1) {};
            \draw[-] (1)--(2)--(3)--(4)--(5) (5)--(6)--(7)--(8);
        \end{tikzpicture}$};
        \node (c) at (0,-2) {$\begin{tikzpicture}
        \node at (3,0) [scale=1, rotate=90]{$\longleftrightarrow$};
        \end{tikzpicture}$};
        \end{tikzpicture}
    \label{rule:chain_rule}
    \end{equation}
Gauging another $\mathbb{Z}_{2}$ flavour subgroup gives rise to a further $\frac{1}{2}$D5-brane at the \emph{other} end of the brane system in \Figref{fig:one_Z2_gauging_dual}. Using the rule of \eqref{rule:chain_rule}, this corresponds to gauging the diagonal $\mathbb{Z}_{2}$ in the other long $DC$-chain of the magnetic theory, which is supported in the field theory by calculating (unrefined) moduli space Hilbert series.
\begin{figure}[h!]
    \centering
    \begin{subfigure}{0.49\textwidth}
    \centering
    \begin{tikzpicture}
        \draw[-](1,1.75)--(4.25,1.75);
        \draw[-](1,0.25)--(4.25,0.25);
        \draw[dotted_brane](-0.5,1)--(0,1);
        \draw[-](0,0)--(0,2);
        \draw[-](0,1)--(1,1);
        \draw[-](1,0)--(1,2);
        \draw[dotted_brane](1,1)--(1.5,1);
        \node (g1) at (1.5,1) [D5] {};
        \draw[dashed_brane](1.5,1)--(2,1);
        \node at (2.75, 0.97) [scale=2]{$\cdots$} {};
        \draw[dotted_brane](3.25,1)--(3.75,1);
        \node (g1) at (3.75,1) [D5] {};
        \draw[dashed_brane](3.75,1)--(4.25,1);
        \draw[-](4.25,0)--(4.25,2);
        \draw [thick,decoration={brace,mirror,raise=0.25cm},decorate] (1.25,1) -- (4,1);
        \node at (2.5,0.5) {$2N_f-1$};
        \node at (2.5,2) {$k$};
    \end{tikzpicture}
    \caption{}
    \label{fig:SQCD_One_Z2_Gauging_A}
    \end{subfigure}
    \begin{subfigure}{0.49\textwidth}
    \centering
    \begin{tikzpicture}
        \node (1) [gauge, label=below:{$\mathfrak{b}_{0}$}] at (0,0) {};
        \node (2) [gauge, label=below:{$\mathfrak{c}_{k}$}] at (1,0) {};
        \node (3) [flavour, label=above:{$\mathfrak{b}_{N_f-1}$}] at (1,1) {};
        \draw[-] (1)--(2)--(3);
    \end{tikzpicture}
    \caption{}
    \label{fig:SQCD_One_Z2_Gauging_B}
    \end{subfigure}
    \caption{The brane system in \Figref{fig:SQCD_One_Z2_Gauging_A} gives rise to the quiver on the right in \Figref{fig:SQCD_One_Z2_Gauging_B}. The leftmost $\frac{1}{2}$NS5-brane interval gives rise to the $\mathbb{Z}_{2}$ gauge node.}
    \label{fig:SQCD_One_Z2_Gauging}
\end{figure}
The conjecture in \eqref{rule:chain_rule} has several simple examples. For $k=1$ and one $\mathbb{Z}_{2}$ gauging the conjecture reduces to the next-to-minimal $\sorm(2\ell+1)$ nilpotent orbit closure quivers given in \cite{Cabrera:2017njm}. For $k=1$ with two $\mathbb{Z}_{2}$ gaugings the conjecture can be tested explicitly using unrefined Hilbert series. For $k=2$ and one $\mathbb{Z}_{2}$ gauging the conjecture precisely replicates the results found in \cite{Cabrera:2017ucb} for the closure of the $[3,2^{2},1]$ orbit of $\mathfrak{so}_{9}$.
\begin{figure}[h!]
\centering
    \begin{tikzpicture}
        \draw[-] (1,0) -- (1,2);
        \draw[-] (2,0) -- (2,2);
        \draw[-] (3,0) -- (3,2);
        \draw[-] (5,0) -- (5,2);
        \draw[-] (6,0) -- (6,2);
        \draw[-] (7,0) -- (7,2);
        \draw[-] (9,0) -- (9,2);
        \draw[-] (10,0) -- (10,2);
        \draw[-] (11,0) -- (11,2);
        \draw[-] (13,0) -- (13,2);
        \draw[-] (14,0) -- (14,2);
        \node (g1) at (0.5,1) [D5] {};
        \draw[dotted_brane] (0,1) -- (0.5,1);
        \draw[dashed_brane] (0.5,1) -- (1,1);
        \draw[dashed_brane] (2,1) -- (3,1);
        \node at (4, 0.97) [scale=2]{$\cdots$} {};
        \draw[dashed_brane] (6,1) -- (6.5,1);
        \draw[dotted_brane] (6.5,1) -- (7,1);
        \node (g2) at (6.5,1) [D5] {};
        \node (g2) at (9.5,1) [D5] {};
        \draw[-] (7,1)--(7.5,1);
        \draw[-] (8.5,1)--(9,1);
        \draw[-] (1,1.4)--(2,1.4);
        \draw[-] (1,0.6)--(2,0.6);
        \draw[-] (2,1.8)--(3,1.8);
        \draw[-] (2,0.2)--(3,0.2);
        \draw[-] (3,1.4)--(3.5,1.4);
        \draw[-] (3,0.6)--(3.5,0.6);
        \node at (8, 0.97) [scale=2]{$\cdots$} {};
        \draw[-] (4.5,1.4)--(5,1.4);
        \draw[-] (4.5,0.6)--(5,0.6);
        \draw[-] (5,1.8)--(6,1.8);
        \draw[-] (5,0.2)--(6,0.2);
        \draw[-] (6,1.4)--(7,1.4);
        \draw[-] (6,0.6)--(7,0.6);
        \draw[-] (7,1.8)--(7.5,1.8);
        \draw[-] (7,0.2)--(7.5,0.2);
        \draw[-] (8.5,1.8)--(9,1.8);
        \draw[-] (8.5,0.2)--(9,0.2);
        \draw[-] (9,1.4)--(10,1.4);
        \draw[-] (9,0.6)--(10,0.6);
        \draw[dotted_brane] (9,1) -- (9.5,1);
        \draw[dashed_brane] (9.5,1) -- (10,1);
        \draw[-] (10,1.8)--(11,1.8);
        \draw[-] (10,0.2)--(11,0.2);
        \draw[-] (11,1.6)--(11.5,1.6);
        \draw[-] (11,0.4)--(11.5,0.4);
        \draw[dashed_brane] (11.3,1) -- (11,1);
        \node at (12, 0.97) [scale=2]{$\cdots$} {};
        \draw[-] (12.5,1.6)--(13,1.6);
        \draw[-] (12.5,0.4)--(13,0.4);
        \draw[dashed_brane] (13,1) -- (14,1);
        \node at (1.5,2) {$1$};
        \node at (2.5,2) {$1$};
        \node at (5.5,2) {$k$};
        \node at (6.5,1.75) {$k$};
        \node at (9.5,1.75) {$k$};
        \node at (10.5,2) {$k$};
    \end{tikzpicture}
    \caption{The brane system in \Figref{fig:one_Z2_gauging_dual} is that of \Figref{fig:SQCD_One_Z2_Gauging_A} under rotation and S-duality. The leftmost NS5-brane in \Figref{fig:SQCD_One_Z2_Gauging_A}, which gives rise to the $\mathbb{Z}_{2}$ gauge symmetry, becomes the leftmost $\frac{1}{2}$D5-brane in \Figref{fig:one_Z2_gauging_dual}. The corresponding quiver is conjectured to be the magnetic theory in the middle row of Table \ref{table:SQCD_with_gaugings_elec_mag}, where the presence of the extra $\frac{1}{2}$D5-brane gives rise to the diagonal $\mathbb{Z}_{2}$ quotient.}
    \label{fig:one_Z2_gauging_dual}
\end{figure}
\subsection{Example}
The rule given in \eqref{rule:chain_rule} can be checked explicitly using the example in \Figref{fig:Diagonal_Gauge_Example_Mirror}. Under \eqref{rule:chain_rule}, the brane system in \Figref{fig:Diagonal_Gauge_Example_A} appears to yield the quiver with the diagonal $\mathbb{Z}_{2}$ gauging given in \Figref{fig:Diagonal_Gauge_Example_B}. \footnote{Note that the monopole formula for \Figref{} deviates from the un-gauged case solely in the dressing factor, following (6.15) in \cite{Cabrera:2017ucb}.} By moving to the magnetic phase of the brane system, given in \Figref{fig:Diagonal_Gauge_Example_Mirror_A}, the corresponding magnetic theory is unambiguously read as \Figref{fig:Diagonal_Gauge_Example_Mirror_B}. Computing the (refined) Higgs branch Hilbert series of \Figref{fig:Diagonal_Gauge_Example_Mirror_A} and the (unrefined) Coulomb branch Hilbert series of \Figref{fig:Diagonal_Gauge_Example_B} shows agreement, as seen in \eqref{eqn:example2BHiggs_unref} and \eqref{eqn:example2BHiggs_ref}. As the diagonal gauging in \Figref{fig:Diagonal_Gauge_Example_B} does not affect the Higgs branch, so does the $\mathbb{Z}_{2}$ flavour subgroup gauging in \Figref{fig:Diagonal_Gauge_Example_Mirror_B} leave the Coulomb branch invariant. As such, $\mathcal{H}(\mathcal{Q}_{\ref{fig:Diagonal_Gauge_Example_B}}) = \mathcal{C}(\mathcal{Q}_{\ref{fig:Diagonal_Gauge_Example_Mirror_B}})$.
\begin{figure}[h!]
    \centering
    \begin{subfigure}{0.49\textwidth}
    \centering
    \begin{tikzpicture}
        \node (g1) at (0.5,1) [D5] {};
        \draw[-] (1,0)--(1,2);
        \draw[-] (2,0)--(2,2);
        \draw[-] (3,0)--(3,2);
        \node (g1) at (3.6,1) [D5] {};
        \node (g1) at (4.2,1) [D5] {};
        \node (g1) at (4.8,1) [D5] {};
        \node (g1) at (5.4,1) [D5] {};
        \draw[-] (6,0)--(6,2);
        \node (g1) at (6.5,1) [D5] {};
        \draw[dashed_brane] (0.5,1) -- (1,1);
        \draw[dashed_brane] (2,1) -- (3,1);
        \draw[dashed_brane] (6,1) -- (6.5,1);
        \draw[dotted_brane] (0,1) -- (0.5,1);
        \draw[dotted_brane] (6.5,1) -- (7,1);
        \draw[-] (1,1.5)--(2,1.5);
        \draw[-] (1,0.5)--(2,0.5);
        \draw[-] (2,1.25)--(3,1.25);
        \draw[-] (2,0.75)--(3,0.75);
        \draw[-] (3,1.5)--(6,1.5);
        \draw[-] (3,0.5)--(6,0.5);
        \draw[-] (3,1.75)--(6,1.75);
        \draw[-] (3,0.25)--(6,0.25);
    \end{tikzpicture}
    \caption{}
    \label{fig:Diagonal_Gauge_Example_A}
    \end{subfigure}
    \begin{subfigure}{0.49\textwidth}
    \centering
    \begin{tikzpicture}
        \node (1) [gauge, label=below:{$[\mathfrak{d}_1$}] at (0,0) {};
        \node (2) [gauge, label=below:{$\mathfrak{c}_1$}] at (1,0) {};
        \node (3) [gauge, label=below:{$\mathfrak{d}_{2}]$}] at (2,0) {};
        \node (4) [flavor, label=above:{$\mathfrak{c}_2$}] at (2,0.75) {};
        \draw[-] (1)--(2)--(3)--(4);
    \end{tikzpicture}
    \caption{}
    \label{fig:Diagonal_Gauge_Example_B}
    \end{subfigure}
    \begin{subfigure}{0.49\textwidth}
    \centering
    \begin{tikzpicture}
        \draw[-] (0,0) -- (0,2);
        \draw[-] (1,0)--(1,2);
        \draw[-] (4,0)--(4,2);
        \draw[-] (5,0)--(5,2);
        \draw[-] (6,0)--(6,2);
        \draw[-] (7,0)--(7,2);
        \draw[-] (0,1)--(1,1);
        \draw[-] (1,1.5)--(4,1.5);
        \draw[-] (1,0.5)--(4,0.5);
        \draw[-] (4,1.25)--(5,1.25);
        \draw[-] (4,0.75)--(5,0.75);
        \draw[-] (4,1)--(5,1);
        \draw[-] (5,1.5)--(6,1.5);
        \draw[-] (5,0.5)--(6,0.5);
        \draw[-] (6,1)--(7,1);
        \draw[-] (1,1.75)--(4,1.75);
        \draw[-] (1,0.25)--(4,0.25);
        \node (g1) at (1.6,1) [D5] {};
        \node (g2) at (2.2,1) [D5] {};
        \node (g3) at (2.8,1) [D5] {};
        \node (g4) at (3.4,1) [D5] {};
        \draw[dotted_brane] (1,1) -- (1.6,1);
        \draw[dotted_brane] (2.2,1) -- (2.8,1);
        \draw[dotted_brane] (3.4,1) -- (4,1);
        \draw[dotted_brane] (5,1) -- (6,1);
        \draw[dashed_brane] (1.6,1) -- (2.2,1);
        \draw[dashed_brane] (2.8,1) -- (3.4,1);
    \end{tikzpicture}
    \caption{}
    \label{fig:Diagonal_Gauge_Example_Mirror_A}
    \end{subfigure}
    \begin{subfigure}{0.49\textwidth}
    \centering
    \begin{tikzpicture}
        \node (1) [gauge, label=below:{$\mathfrak{b}_0$}] at (0,0) {};
        \node (2) [gauge, label=below:{$\mathfrak{c}_2$}] at (1,0) {};
        \node (3) [gauge, label=below:{$\mathfrak{b}_{1}$}] at (2,0) {};
        \node (4) [gauge, label=below:{$\mathfrak{c}_1$}] at (3,0) {};
        \node (5) [gauge, label=below:{$\mathfrak{b}_0$}] at (4,0) {};
        \node (6) [flavor, label=above:{$\mathfrak{d}_2$}] at (1,0.75) {};
        \draw[-] (1)--(2)--(6)--(2)--(3)--(4)--(5);
    \end{tikzpicture}
    \caption{}
    \label{fig:Diagonal_Gauge_Example_Mirror_B}
    \end{subfigure}
    \caption{The brane system in \Figref{fig:Diagonal_Gauge_Example_A} yields the quiver \Figref{fig:Diagonal_Gauge_Example_B} under the application of the rule given in \eqref{rule:chain_rule}. The $\frac{1}{2}$D5-branes at each end of the maximal chain gauge the diagonal $\mathbb{Z}_{2}$ symmetry on the magnetic lattice. The brane system in \Figref{fig:Diagonal_Gauge_Example_Mirror_A} results from performing an S-duality transformation on that of \Figref{fig:Diagonal_Gauge_Example_A}. The associated theory, given in \Figref{fig:Diagonal_Gauge_Example_B}, is read unambiguously.}
    \label{fig:Diagonal_Gauge_Example_Mirror}
\end{figure}
\begin{equation}
    \text{HS}_{\mathcal{H}}(\mathcal{Q}_{\ref{fig:Diagonal_Gauge_Example_Mirror_B}}) =\begin{aligned} &\pe\left[([2;0]+[0;2])t^2+([2;2]-[2;0]-[2;0]-1)t^4\right]\\&\times(1+t^4)(1+([2;0]+[0;2])t^4+t^8)\end{aligned}
    \label{eqn:example2BHiggs_ref}
    \end{equation}
    \begin{equation}
    \text{HS}_{\mathcal{H}}(\mathcal{Q}_{\ref{fig:Diagonal_Gauge_Example_Mirror_B}})\rvert_{a,b\rightarrow1} = \frac{(1+t^4)(1+6t^4+t^8)}{(1-t^2)^{6}(1-t^4)^{2}} = \text{HS}_{\mathcal{C}}(\mathcal{Q}_{\ref{fig:Diagonal_Gauge_Example_B}})
\label{eqn:example2BHiggs_unref}
\end{equation}
Hence, the moduli space calculations support the interpretation of \Figref{} and \Figref{} as a 3d mirror pair -- note that without the diagonal $\mathbb{Z}_{2}$ gauging the Coulomb branch of \Figref{} would not match the Higgs branch of \Figref{}. The interplay between diagonal $\mathbb{Z}_{2}$ quotients on magnetic lattices and $\mathbb{Z}_{2}$ flavour gaugings appears to be generic for this class of framed orthosympelctic theories.
\begin{landscape}
\begin{table}[h!]
\ra{1.5}
    \centering
    \begin{tabular}{ccc}
    \toprule
         Electric Theory & Magnetic Theory \\ \midrule
        \begin{tikzpicture}
        \node (1) [flavor, label=above:{$\mathfrak{d}_{2k+1+N}$}] at (0,1) {};
        \node (2) [gauge, label=below:{$\mathfrak{c}_{k}$}] at (0,0) {};
        \draw (1)--(2);
        \end{tikzpicture} & \begin{tikzpicture}
    \node (1) [gauge, label=below:{$\mathfrak{d}_1$}] at (0,0) {};
    \node (2) [gauge, label=below:{$\mathfrak{c}_1$}] at (1,0) {};
    \node (3) [gauge, label=below:{$\mathfrak{d}_{k}$}] at (3,0) {};
    \node (4) [gauge, label=below:{$\mathfrak{c}_{k}$}] at (4,0) {};
    \node (6) [flavor, label=above:{$\mathfrak{b}_0$}] at (4,1) {};
    \node (7) [gauge, label=below:{$\mathfrak{b}_{k}$}] at (5,0) {};
    \node (8) [gauge, label=below:{$\mathfrak{b}_k$}] at (7,0) {};
    \node (9) [gauge, label=below:{$\mathfrak{c}_k$}] at (8,0) {};
    \node (10) [flavor, label=above:{$\mathfrak{b}_0$}] at (8,1) {};
    \node (11) [gauge, label=below:{$\mathfrak{d}_k$}] at (9,0) {};
    \node (12) [gauge, label=below:{$\mathfrak{c}_1$}] at (11,0) {};
    \node (13) [gauge, label=below:{$\mathfrak{d}_1$}] at (12,0) {};
    \draw (1) -- (2);
    \draw (3) -- (4) -- (6);
    \draw (2) -- (1.5,0);
    \draw (3) -- (2.5,0);
    \draw (4)--(7)--(5.5,0);
    \draw (6.5,0)--(8)--(9);
    \draw (13)--(12)--(10.5,0);
    \draw (9.5,0)--(11)--(9)--(10)--(9);
    \node at (2.04, -0.03) [scale=2]{$\cdots$} {};
    \node at (6.04, -0.03) [scale=2]{$\cdots$} {};
    \node at (10.04, -0.03) [scale=2]{$\cdots$} {};
    \draw [thick,decoration={brace,mirror,raise=0cm},decorate] (4,-0.75) -- (8,-0.75);
    \node at (6,-1) {$2N+1$};
    \end{tikzpicture}\\ \midrule
        \begin{tikzpicture}
        \node (1) [flavor, label=above:{$\mathfrak{b}_{2k+N}$}] at (0,1) {};
        \node (2) [gauge, label=below:{$\mathfrak{c}_{k}$}] at (0,0) {};
        \node (3) [gauge, label=below:{$\mathfrak{b}_{0}$}] at (1,0) {};
        \draw (1)--(2)--(3);
        \end{tikzpicture} &  \begin{tikzpicture}
    \node (1) [gauge, label=below:{$[\mathfrak{d}_1$}] at (0,0) {};
    \node (2) [gauge, label=below:{$\mathfrak{c}_1$}] at (1,0) {};
    \node (3) [gauge, label=below:{$\mathfrak{d}_{k}]$}] at (3,0) {};
    \node (4) [gauge, label=below:{$\mathfrak{c}_{k}$}] at (4,0) {};
    \node (6) [flavor, label=above:{$\mathfrak{b}_0$}] at (4,1) {};
    \node (7) [gauge, label=below:{$\mathfrak{b}_{k}$}] at (5,0) {};
    \node (8) [gauge, label=below:{$\mathfrak{b}_k$}] at (7,0) {};
    \node (9) [gauge, label=below:{$\mathfrak{c}_k$}] at (8,0) {};
    \node (10) [flavor, label=above:{$\mathfrak{b}_0$}] at (8,1) {};
    \node (11) [gauge, label=below:{$\mathfrak{d}_k$}] at (9,0) {};
    \node (12) [gauge, label=below:{$\mathfrak{c}_1$}] at (11,0) {};
    \node (13) [gauge, label=below:{$\mathfrak{d}_1$}] at (12,0) {};
    \draw (1) -- (2);
    \draw (3) -- (4) -- (6);
    \draw (2) -- (1.5,0);
    \draw (3) -- (2.5,0);
    \draw (4)--(7)--(5.5,0);
    \draw (6.5,0)--(8)--(9);
    \draw (13)--(12)--(10.5,0);
    \draw (9.5,0)--(11)--(9)--(10)--(9);
    \node at (2.04, -0.03) [scale=2]{$\cdots$} {};
    \node at (6.04, -0.03) [scale=2]{$\cdots$} {};
    \node at (10.04, -0.03) [scale=2]{$\cdots$} {};
    \draw [thick,decoration={brace,mirror,raise=0cm},decorate] (4,-0.75) -- (8,-0.75);
    \node at (6,-1) {$2N+1$};
    \end{tikzpicture}\\ \midrule
        \begin{tikzpicture}
        \node (1) [flavor, label=above:{$\mathfrak{d}_{2k+N}$}] at (0,1) {};
        \node (2) [gauge, label=below:{$\mathfrak{c}_{k}$}] at (0,0) {};
        \node (3) [gauge, label=right:{$\mathfrak{b}_{0}$}] at (1,0) {};
        \node (4) [gauge, label=right:{$\mathfrak{b}_{0}$}] at (1,0.75) {};
        \draw (1)--(2)--(3)--(2)--(4);
        \end{tikzpicture} & \begin{tikzpicture}
    \node (1) [gauge, label=below:{$[\mathfrak{d}_1$}] at (0,0) {};
    \node (2) [gauge, label=below:{$\mathfrak{c}_1$}] at (1,0) {};
    \node (3) [gauge, label=below:{$\mathfrak{d}_{k}]$}] at (3,0) {};
    \node (4) [gauge, label=below:{$\mathfrak{c}_{k}$}] at (4,0) {};
    \node (6) [flavor, label=above:{$\mathfrak{b}_0$}] at (4,1) {};
    \node (7) [gauge, label=below:{$\mathfrak{b}_{k}$}] at (5,0) {};
    \node (8) [gauge, label=below:{$\mathfrak{b}_k$}] at (7,0) {};
    \node (9) [gauge, label=below:{$\mathfrak{c}_k$}] at (8,0) {};
    \node (10) [flavor, label=above:{$\mathfrak{b}_0$}] at (8,1) {};
    \node (11) [gauge, label=below:{$[\mathfrak{d}_k$}] at (9,0) {};
    \node (12) [gauge, label=below:{$\mathfrak{c}_1$}] at (11,0) {};
    \node (13) [gauge, label=below:{$\mathfrak{d}_1$]}] at (12,0) {};
    \draw (1) -- (2);
    \draw (3) -- (4) -- (6);
    \draw (2) -- (1.5,0);
    \draw (3) -- (2.5,0);
    \draw (4)--(7)--(5.5,0);
    \draw (6.5,0)--(8)--(9);
    \draw (13)--(12)--(10.5,0);
    \draw (9.5,0)--(11)--(9)--(10)--(9);
    \node at (2.04, -0.03) [scale=2]{$\cdots$} {};
    \node at (6.04, -0.03) [scale=2]{$\cdots$} {};
    \node at (10.04, -0.03) [scale=2]{$\cdots$} {};
    \draw [thick,decoration={brace,mirror,raise=0cm},decorate] (4,-0.75) -- (8,-0.75);
    \node at (6,-1) {$2N+1$};
    \end{tikzpicture}\\
         \bottomrule
    \end{tabular}
    \caption{The electric-magnetic pairs derived from brane systems for $\sprm(k)$ SQCD with on or two $\mathbb{Z}_{2}$ gaugings. The gauging in the magnetic theory appears as a diagonal quotient on the magnetic lattice.}
    \label{table:SQCD_with_gaugings_elec_mag}
\end{table}
\end{landscape}
\section{Outlook}
This paper introduced several conjectures regarding the interpretation of discrete gaugings in brane systems involving D3-/D5-/NS5-branes with $O3$ orientifold planes and identities on the Coulomb branch from Higgsings within a special piece, supported using 3d mirror symmetry and Hilbert series computations.

It is also interesting to consider the fate of the Coulomb branch identity in Section \ref{quiv:identity:general} under 3d mirror symmetry. It appears that these identities lead to a map on the Higgs branch of a maximal $BC$-chain that decreases the rank of each node of $B$-type. Calculations show that this map keeps the Higgs branch invariant, despite manifest incomplete Higgsing. Similarly, calculations show that the diagonal $\mathbb{Z}_{2}$ gauging of a maximal $DC$-chain does not change the Higgs branch, which agrees with intuition from its electric counterpart -- it would be useful to prove these statements.

It is also unclear as to why the Higgsing in Section \ref{sec:As_A_Higgsing} requires \emph{all} $\frac{1}{2}$D3-branes created under the splitting of a D5-brane to be removed. Removing only \emph{some} of the red branes in \Figref{fig:generic_identity_brane_interp_B} leads to a brane system whose corresponding quiver theory cannot be read using current techniques.

It is possible to conjecture rules for framed orthosymplectic subtraction and perform rudimentary examples in the presence of the $\mathbb{Z}_{2}$ gaugings identified in this work. However, fundamental problems regarding the interpretation of D3-/D5-/NS5-brane systems with $O3$ orientifold planes circumscribe the set of identifiable transitions to those of type $ADE$ with some exceptions \cite{Cabrera:2017njm}. Moreover, the brane system interpretation of the two cones in the Higgs branch of $\sprm(k)$ gauge theory with $2k$ flavours is also unclear \cite{Ferlito:2016grh, Bourget:2023cgs}. It would be helpful to settle these challenges in future work.

Given the various computational constraints imposed by orthosymplectic quivers, much of the topic remains unclear. In the first case, the identifications in \eqref{rule:O_type_gauge} and \eqref{rule:chain_rule} immediately give rise to the question of a string theory interpretation. A satisfactory answer may currently remain out of reach; it is unclear for instance whether this bears any relation to the introduction of spinor matter in \cite{Zafrir:2015ftn}.
\acknowledgments
We thank Rudolph Kalveks, Guhesh Kumaran, Hiraku Nakajima, Michael Finkelberg, Noppadol Mekareeya and William Harding for helpful discussions. The work of SB and AH is partially supported by STFC Consolidated Grants ST/T000791/1 and ST/X000575/1. SB is supported by the STFC DTP research studentship grant ST/Y509231/1.
\appendix
\section{Evidence for Identities}
\label{app:proof_of_identities}
This appendix contains evidence for the identities \eqref{quiv:identity:m=n=0}, \eqref{quiv:identity:m=n=1} and \eqref{quiv:identity:m=n_general} using the monopole formula to calculate unrefined Hilbert series. Identical unrefined Hilbert series are not sufficient for proving an equivalence of moduli spaces - at best they provide a strong indication of a match. In analogy to that given for the $BC$ chain in Section \ref{sec:BC_Chains}, the proofs here proceed by recognising that the magnetic lattices and dressing factors are unchanged on either side of \eqref{quiv:identity:m=n=0}, \eqref{quiv:identity:m=n=1} and \eqref{quiv:identity:m=n_general}. Hence it suffices to check that the conformal dimensions of the two quivers on each side of a given identity are equal, calculated explicitly below.
\begin{figure}[h!]
\centering
\begin{subfigure}{0.49\textwidth}
    \centering
    \scalebox{1}{\begin{tikzpicture}
        \node (1) [flavor, label=above:{$\mathfrak{c}_{k_2}$}] at (0,1) {};
        \node (2) [gauge, label=below:{$\mathfrak{b}_{k}$}] at (0,0) {};
        \node (3) [gauge, label=below:{$\mathfrak{c}_{k_3}$}] at (1,0) {};
        \node (4) [gauge, label=below:{$\mathfrak{c}_{k_1}$}] at (-1,0) {};
        \draw (1)--(2)--(3)--(2)--(4);
        \node at (-1,-1) {$m_{i}^{1}$};
        \node at (0,-1) {$m_{j}^{2}$};
        \node at (1,-1) {$m_{l}^{3}$};
        \node at (-3,1) {$i=1,\cdots,k_1$};
        \node at (-3,0) {$j=1,\cdots,k$};
        \node at (-3,-1) {$l=1,\cdots,k_3$};
        \end{tikzpicture}}
        \caption{}
        \label{fig:iden_1_annotated1}
    \end{subfigure}
    \begin{subfigure}{0.49\textwidth}
    \centering
    \scalebox{1}{\begin{tikzpicture}
       \node (1) [gauge, label=below:{$\orm(2k)$}] at (0,0) {};
        \node (2) [gauge, label=below:{$\mathfrak{c}_{k_3}$}] at (1,0) {};
        \node (3) [flavor, label=above:{$\mathfrak{b}_0$}] at (1,1) {};
        \node (4) [flavor, label=above:{$\mathfrak{c}_{k_2-1}$}] at (0,1) {};
        \node (5) [gauge, label=below:{$\mathfrak{c}_{k_1}$}] at (-1,0) {};
        \node (6) [flavor, label=above:{$\mathfrak{b}_0$}] at (-1,1) {};
        \draw (4)--(1)--(2)--(3)--(2)--(1)--(5)--(6);
        \node at (-1,-1) {$g_{i}^{1}$};
        \node at (0,-1) {$g_{j}^{2}$};
        \node at (1,-1) {$g_{l}^{3}$};
        \node at (3,1) {$i=1,\cdots,k_1$};
        \node at (3,0) {$j=1,\cdots,k$};
        \node at (3,-1) {$l=1,\cdots,k_3$};
        \end{tikzpicture}}
        \caption{}
        \label{fig:iden1_annotated2}
    \end{subfigure}
\caption{The identity given in \eqref{quiv:identity:m=n=0} with magnetic charges assigned to each gauge node.}
\label{fig:iden1_Annotated}
\end{figure}
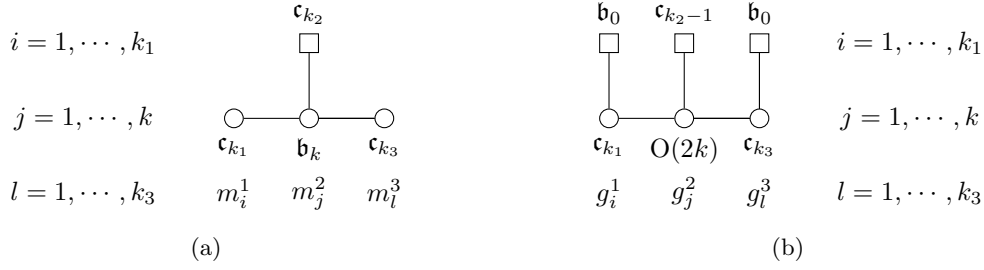
First consider the identity given in \eqref{quiv:identity:m=n=0}, rewritten with its magnetic charges in \Figref{fig:iden1_Annotated}. The conformal dimension $\Delta_{\text{L}}$ of \Figref{fig:iden_1_annotated1} is given in \eqref{eqn:appnA1}, where $\Sigma_{\mathfrak{c}_{k_1}}$ and $\Sigma_{\mathfrak{c}_{k_3}}$ denote the vectormultiplet contributions from the $\sprm(k_1)$ and $\sprm(k_3)$ nodes and $\Sigma_{m}$ collects terms of the form $\left|m^{\alpha} \pm m^{\beta}\right|$ ($\alpha, \beta$ = 1,2,3) common to both $\Delta_{\text{L}}$ and $\Delta_{\text{R}}$.
\begin{align}
    \Delta_{\text{L}}^{\text{Hyp}} =& \; \frac{1}{2}\sum_{i,j} \left( \left|m^{1}_{i}-m^{2}_{j}\right| + \left|m^{1}_{i}+m^{2}_{j}\right|\right) + \frac{1}{2}\sum_{i}\left|m^{1}_{i}\right|+\frac{1}{2}\sum_{j,l} \left( \left|m^{2}_{j}-m^{3}_{l}\right| + \left|m^{2}_{j}+m^{3}_{l}\right|\right)\\&+ \frac{1}{2}\sum_{l}\left|m^{3}_{i}\right| + k_2\sum_{j}\left|m^{2}_{j}\right|\\
    \Delta_{\text{L}}^{\text{Vec}} =& \; \Sigma_{\mathfrak{c}_{k_1}}+\Sigma_{\mathfrak{c}_{k_3}} - \sum_{a<b}\left(\left|m^{2}_{a}+m^{2}_{b}\right|+\left|m^{2}_{a}-m^{2}_{b}\right|\right) - \sum_{j}\left|m^{2}_{j}\right|\\
    \Delta_{\text{L}} =&  \; \Delta_{\text{L}}^{\text{Hyp}}+\Delta_{\text{L}}^{\text{Vec}} = \; \Sigma_{m} +\frac{1}{2}\sum_{i}\left|m^{1}_{i}\right|+\frac{1}{2}\sum_{l}\left|m^{3}_{l}\right|+(k_2-1)\sum_{j}\left|m^{2}_{j}\right|
    \label{eqn:appnA1}
\end{align}
The conformal dimension $\Delta_{\text{R}}$ of \Figref{fig:iden1_annotated2} is similarly computed in \eqref{eqn:appnA2}.
\begin{align}
    \Delta_{\text{R}}^{\text{Hyp}} =& \frac{1}{2}\sum_{i,j} \left( \left|g^{1}_{i}-g^{2}_{j}\right| + \left|g^{1}_{i}+g^{2}_{j}\right|\right)+\frac{1}{2}\sum_{j,l} \left( \left|g^{2}_{j}-g^{3}_{l}\right| + \left|g^{2}_{j}+g^{3}_{l}\right|\right)+ \frac{1}{2}\sum_{i}\left|g^{1}_{i}\right|\\&+ \frac{1}{2}\sum_{l}\left|g^{3}_{i}\right| + (k_2-1)\sum_{j}\left|g^{2}_{j}\right|\\
    \Delta_{\text{R}}^{\text{Vec}} =& \; \Sigma_{\mathfrak{c}_{k_1}}+\Sigma_{\mathfrak{c}_{k_3}} - \sum_{a<b}\left(\left|g^{2}_{a}+g^{2}_{b}\right|+\left|g^{2}_{a}-g^{2}_{b}\right|\right)\\
    \Delta_{\text{R}} =& \Delta_{\text{R}}^{\text{Hyp}}+\Delta_{\text{R}}^{\text{Vec}} = \; \Sigma_{g} +\frac{1}{2}\sum_{i}\left|g^{1}_{i}\right|+\frac{1}{2}\sum_{l}\left|g^{3}_{l}\right|+(k_2-1)\sum_{j}\left|g^{2}_{j}\right|
    \label{eqn:appnA2}
\end{align}
Since $m^{1,2,3}$ and $g^{1,2,3}$ are summed over the same lattice, the conformal dimension contributions of \Figref{fig:iden_1_annotated1} and \Figref{fig:iden1_annotated2} are identical.

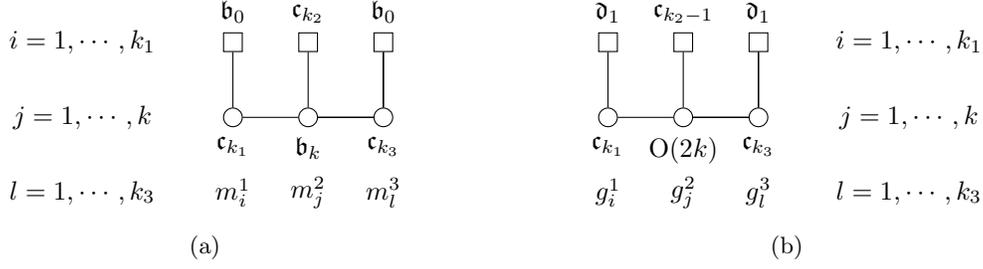
\begin{figure}[h!]
\centering
\begin{subfigure}{0.49\textwidth}
    \centering
    \scalebox{1}{\begin{tikzpicture}
        \node (1) [flavor, label=above:{$\mathfrak{c}_{k_2}$}] at (0,1) {};
        \node (2) [gauge, label=below:{$\mathfrak{b}_{k}$}] at (0,0) {};
        \node (3) [gauge, label=below:{$\mathfrak{c}_{k_3}$}] at (1,0) {};
        \node (4) [gauge, label=below:{$\mathfrak{c}_{k_1}$}] at (-1,0) {};
        \node (5) [flavor, label=above:{$\mathfrak{b}_{0}$}] at (-1,1) {};
        \node (6) [flavor, label=above:{$\mathfrak{b}_{0}$}] at (1,1) {};
        \draw (1)--(2)--(3)--(6)--(3)--(2)--(4)--(5);
        \node at (-1,-1) {$m_{i}^{1}$};
        \node at (0,-1) {$m_{j}^{2}$};
        \node at (1,-1) {$m_{l}^{3}$};
        \node at (-3,1) {$i=1,\cdots,k_1$};
        \node at (-3,0) {$j=1,\cdots,k$};
        \node at (-3,-1) {$l=1,\cdots,k_3$};
        \end{tikzpicture}}
        \caption{}
        \label{fig:iden_2_annotated1}
    \end{subfigure}
    \begin{subfigure}{0.49\textwidth}
    \centering
    \scalebox{1}{\begin{tikzpicture}
       \node (1) [gauge, label=below:{$\orm(2k)$}] at (0,0) {};
        \node (2) [gauge, label=below:{$\mathfrak{c}_{k_3}$}] at (1,0) {};
        \node (3) [flavor, label=above:{$\mathfrak{d}_1$}] at (1,1) {};
        \node (4) [flavor, label=above:{$\mathfrak{c}_{k_2-1}$}] at (0,1) {};
        \node (5) [gauge, label=below:{$\mathfrak{c}_{k_1}$}] at (-1,0) {};
        \node (6) [flavor, label=above:{$\mathfrak{d}_1$}] at (-1,1) {};
        \draw (4)--(1)--(2)--(3)--(2)--(1)--(5)--(6);
        \node at (-1,-1) {$g_{i}^{1}$};
        \node at (0,-1) {$g_{j}^{2}$};
        \node at (1,-1) {$g_{l}^{3}$};
        \node at (3,1) {$i=1,\cdots,k_1$};
        \node at (3,0) {$j=1,\cdots,k$};
        \node at (3,-1) {$l=1,\cdots,k_3$};
        \end{tikzpicture}}
        \caption{}
        \label{fig:iden_2_annotated2}
    \end{subfigure}
\caption{The identity given in \eqref{quiv:identity:m=n=1} with magnetic charges assigned to each gauge node.}
\label{fig:iden2_Annotated}
\end{figure}

The identity given in \eqref{quiv:identity:m=n=1}, rewritten in \Figref{fig:iden2_Annotated} alongside its magnetic charges, is similarly supported by a calculation using the unrefined Hilbert series.
The conformal dimension for the quiver in \Figref{fig:iden_2_annotated1} is given in \eqref{eqn:appnA3}, in which
\begin{align}
    \Delta_{\text{L}}^{\text{Hyp}} =& \; \frac{1}{2}\sum_{i,j} \left( \left|m^{1}_{i}-m^{2}_{j}\right| + \left|m^{1}_{i}+m^{2}_{j}\right|\right) + \sum_{i}\left|m^{1}_{i}\right|+\frac{1}{2}\sum_{j,l} \left( \left|m^{2}_{j}-m^{3}_{l}\right| + \left|m^{2}_{j}+m^{3}_{l}\right|\right)\\&+ \sum_{l}\left|m^{3}_{i}\right| + k_2\sum_{j}\left|m^{2}_{j}\right|\\
    \Delta_{\text{L}}^{\text{Vec}} =& \; \Sigma_{\mathfrak{c}_{k_1}}+\Sigma_{\mathfrak{c}_{k_3}} - \sum_{a<b}\left(\left|m^{2}_{a}+m^{2}_{b}\right|+\left|m^{2}_{a}-m^{2}_{b}\right|\right) - \sum_{j}\left|m^{2}_{j}\right|\\
    \Delta_{\text{L}} =&  \; \Delta_{\text{L}}^{\text{Hyp}}+\Delta_{\text{L}}^{\text{Vec}} = \; \Sigma_{m} +\sum_{i}\left|m^{1}_{i}\right|+\sum_{l}\left|m^{3}_{l}\right|+(k_2-1)\sum_{j}\left|m^{2}_{j}\right|
    \label{eqn:appnA3}
\end{align}
The conformal dimension for the quiver in \Figref{fig:iden_2_annotated2} is given in \eqref{eqn:appnA4}
\begin{align}
    \Delta_{\text{R}}^{\text{Hyp}} =& \frac{1}{2}\sum_{i,j} \left( \left|g^{1}_{i}-g^{2}_{j}\right| + \left|g^{1}_{i}+g^{2}_{j}\right|\right)+\frac{1}{2}\sum_{j,l} \left( \left|g^{2}_{j}-g^{3}_{l}\right| + \left|g^{2}_{j}+g^{3}_{l}\right|\right)+ \sum_{i}\left|g^{1}_{i}\right|\\&+ \sum_{l}\left|g^{3}_{l}\right| + (k_2-1)\sum_{j}\left|g^{2}_{j}\right|\\
    \Delta_{\text{R}}^{\text{Vec}} =& \; \Sigma_{\mathfrak{c}_{k_1}}+\Sigma_{\mathfrak{c}_{k_3}} - \sum_{a<b}\left(\left|g^{2}_{a}+g^{2}_{b}\right|+\left|g^{2}_{a}-g^{2}_{b}\right|\right)\\
    \Delta_{\text{R}} =& \Delta_{\text{R}}^{\text{Hyp}}+\Delta_{\text{R}}^{\text{Vec}} = \; \Sigma_{g} +\sum_{i}\left|g^{1}_{i}\right|+\sum_{l}\left|g^{3}_{l}\right|+(k_2-1)\sum_{j}\left|g^{2}_{j}\right|
    \label{eqn:appnA4}
\end{align}
It's clear by now that for general $m,n$ in \eqref{quiv:identity:m=n_general} the calculation proceeds in almost exactly the same way.
\begin{figure}[H]
\centering
\begin{subfigure}{0.49\textwidth}
    \centering
    \scalebox{1}{\begin{tikzpicture}
        \node (1) [flavor, label=above:{$\mathfrak{c}_{k_2}$}] at (0,1) {};
        \node (2) [gauge, label=below:{$\mathfrak{b}_{k}$}] at (0,0) {};
        \node (3) [gauge, label=below:{$\mathfrak{c}_{k_3}$}] at (1,0) {};
        \node (4) [gauge, label=below:{$\mathfrak{c}_{k_1}$}] at (-1,0) {};
        \node (5) [flavor, label=above:{$\mathfrak{d}_{p}$}] at (-1,1) {};
        \node (6) [flavor, label=above:{$\mathfrak{b}_{q}$}] at (1,1) {};
        \draw (1)--(2)--(3)--(6)--(3)--(2)--(4)--(5);
        \node at (-1,-1) {$m_{i}^{1}$};
        \node at (0,-1) {$m_{j}^{2}$};
        \node at (1,-1) {$m_{l}^{3}$};
        \node at (-3,1) {$i=1,\cdots,k_1$};
        \node at (-3,0) {$j=1,\cdots,k$};
        \node at (-3,-1) {$l=1,\cdots,k_3$};
        \end{tikzpicture}}
        \caption{}
        \label{fig:iden_3_annotated1}
    \end{subfigure}
    \begin{subfigure}{0.49\textwidth}
    \centering
    \scalebox{1}{\begin{tikzpicture}
       \node (1) [gauge, label=below:{$\orm(2k)$}] at (0,0) {};
        \node (2) [gauge, label=below:{$\mathfrak{c}_{k_3}$}] at (1,0) {};
        \node (3) [flavor, label=above:{$\mathfrak{d}_{q+1}$}] at (1,1) {};
        \node (4) [flavor, label=above:{$\mathfrak{c}_{k_2-1}$}] at (0,1) {};
        \node (5) [gauge, label=below:{$\mathfrak{c}_{k_1}$}] at (-1,0) {};
        \node (6) [flavor, label=above:{$\mathfrak{b}_p$}] at (-1,1) {};
        \draw (4)--(1)--(2)--(3)--(2)--(1)--(5)--(6);
        \node at (-1,-1) {$g_{i}^{1}$};
        \node at (0,-1) {$g_{j}^{2}$};
        \node at (1,-1) {$g_{l}^{3}$};
        \node at (3,1) {$i=1,\cdots,k_1$};
        \node at (3,0) {$j=1,\cdots,k$};
        \node at (3,-1) {$l=1,\cdots,k_3$};
        \end{tikzpicture}}
        \caption{}
        \label{fig:iden3_annotated2}
    \end{subfigure}
\caption{The identity given in \eqref{quiv:identity:m=n_general} with magnetic charges assigned to each gauge node.}
\label{fig:iden3_Annotated}
\end{figure}
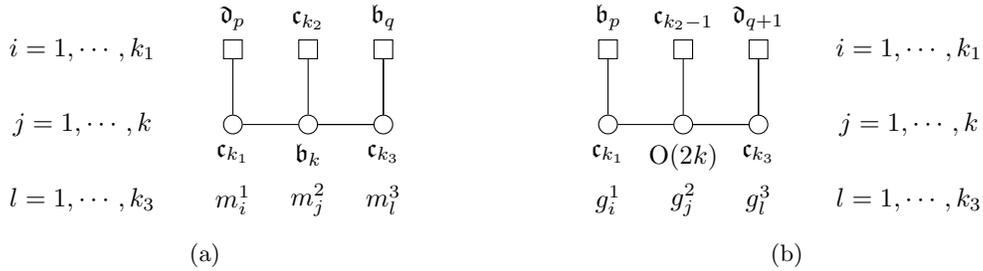
The conformal dimension for the quiver in \Figref{fig:iden_3_annotated1}, $\Delta_{\text{L}}$, is given in \eqref{eqn:appnA5}.
\begin{align}
    \Delta_{\text{L}}^{\text{Hyp}} =& \; \frac{1}{2}\sum_{i,j} \left( \left|m^{1}_{i}-m^{2}_{j}\right| + \left|m^{1}_{i}+m^{2}_{j}\right|\right) + p\sum_{i}\left|m^{1}_{i}\right|+\frac{1}{2}\sum_{j,q} \left( \left|m^{2}_{j}-m^{3}_{q}\right| + \left|m^{2}_{j}+m^{3}_{q}\right|\right)\\&+ l\sum_{q}\left|m^{3}_{q}\right| + \frac{1}{2}\sum_{q}\left|m^{3} _{q}\right| + \frac{1}{2}\sum_{q}\left|m^{3} _{q}\right| + k_2\sum_{j}\left|m^{2}_{j}\right| + \frac{1}{2}\sum_{i}\left|m^{1}_{i}\right|\\
    \Delta_{\text{L}}^{\text{Vec}} =& \; \Sigma_{\mathfrak{c}_{k_1}}+\Sigma_{\mathfrak{c}_{k_3}} - \sum_{a<b}\left(\left|m^{2}_{a}+m^{2}_{b}\right|+\left|m^{2}_{a}-m^{2}_{b}\right|\right) - \sum_{j}\left|m^{2}_{j}\right| - \sum_{j}\left|m^{2}_{j}\right|\\
    \Delta_{\text{L}} =&  \; \Delta_{\text{L}}^{\text{Hyp}}+\Delta_{\text{L}}^{\text{Vec}} = \; \Sigma_{m} + (p+\frac{1}{2})\sum_{i}\left|m^{1}_{i}\right|+(l+1)\sum_{q}\left|m^{3}_{q}\right|+(k_2-1)\sum_{j}\left|m^{2}_{j}\right|
    \label{eqn:appnA5}
\end{align}
The conformal dimension for the quiver in \Figref{fig:iden3_annotated2} is similarly given in \eqref{eqn:appnA6}.
\begin{align}
    \Delta_{\text{R}}^{\text{Hyp}} =& \frac{1}{2}\sum_{i,j} \left( \left|g^{1}_{i}-g^{2}_{j}\right| + \left|g^{1}_{i}+g^{2}_{j}\right|\right)+\frac{1}{2}\sum_{j,q} \left( \left|g^{2}_{j}-g^{3}_{q}\right| + \left|g^{2}_{j}+g^{3}_{q}\right|\right)+ p\sum_{i}\left|g^{1}_{i}\right|\\&+ (l+1)\sum_{q}\left|g^{3}_{q}\right| + (k_2-1)\sum_{j}\left|g^{2}_{j}\right|+\frac{1}{2}\sum_{i}\left|g^{1}_{i}\right|\\
    \Delta_{\text{R}}^{\text{Vec}} =& \; \Sigma_{\mathfrak{c}_{k_1}}+\Sigma_{\mathfrak{c}_{k_3}} - \sum_{a<b}\left(\left|g^{2}_{a}+g^{2}_{b}\right|+\left|g^{2}_{a}-g^{2}_{b}\right|\right)\\
    \Delta_{\text{R}} =& \Delta_{\text{R}}^{\text{Hyp}}+\Delta_{\text{R}}^{\text{Vec}} = \; \Sigma_{g} +(p+\frac{1}{2})\sum_{i}\left|g^{1}_{i}\right|+(l+1)\sum_{q}\left|g^{3}_{q}\right|+(k_2-1)\sum_{j}\left|g^{2}_{j}\right|
    \label{eqn:appnA6}
\end{align}
Again, since the $m$ and $g$ magnetic charges are summed over the same lattice, the two conformal dimensions contribute identically to their respective sums.

\newpage
\section{Hilbert Series for $BC$ Chains}
\label{app:BC_Chain_HS}
This appendix contains the full unrefined Hilbert series for several of the $BC$-chain quivers $\mathcal{Q}_{\ref{quiv:appendix_BC_chain}}$.
\begin{equation}
    \begin{tikzpicture}
    \node (1) [gauge, label=below:{$\mathfrak{c}_k$}] at (0,0) {};
    \node (2) [gauge, label=below:{$\mathfrak{b}_k$}] at (1,0) {};
    \node (3) [gauge, label=below:{$\mathfrak{b}_k$}] at (4,0) {};
    \node (4) [gauge, label=below:{$\mathfrak{c}_k$}] at (5,0) {};
    \node (5) [flavor, label=above:{$\mathfrak{b}_k$}] at (0,1) {};
    \node (6) [flavor, label=above:{$\mathfrak{b}_k$}] at (5,1) {};
    \draw (5) -- (1) -- (2);
    \draw (3) -- (4) -- (6);
    \draw (2) -- (2,0);
    \draw (3) -- (3,0);
    \node at (2.54, -0.03) [scale=2]{$\cdots$} {};
    \draw [thick,decoration={brace,mirror,raise=0.75cm},decorate] (-0.25,0) -- (5.25,0);
    \node at (2.5,-1.25) {$2N+1$};
    \end{tikzpicture}
\label{quiv:appendix_BC_chain}
\end{equation}
\begin{table}[h]
\ra{1.5}
    \centering
    \begin{tabular}{cccc}
    \toprule
         $k$ & $N$  &  Hilbert Series & PL\\ \midrule
         
         1 & 1 & $\frac{1+3t^{2}+11t^{4}+10t^{6}+11t^{8}+3t^{10}+t^{12}}{(1-t^{4})^{3}(1-t^{2})^{3}}$ & $6t^{2}+8t^{4}-15t^{6}-4t^{8}+\cdots$
         \\ \midrule
         & 2 & $\frac{1+10t^{2}+55t^{4}+150t^{6}+288t^{8}+336t^{10}+288t^{12}+150t^{14}+55t^{16}+10t^{18}+t^{20}}{(1-t^{4})^{5}(1-t^{2})^{5}}$ & $15t^2+5t^4-70t^6+273t^8+\cdots$\\ \midrule
         & 3 & $\frac{1+21t^{2}+189t^{4}+931t^{6}+3003t^{8}+6615t^{10}+10567t^{12}+12258t^{14}+\cdots+t^{28}}{(1-t^{4})^{7}(1-t^{2})^{7}}$ & $28t^2-35t^4+42t^6+336t^8+\cdots$\\
         \midrule
         2 & 1 & $\frac{1+6t^{4}+6t^{6}+26t^{8}+15t^{10}+76t^{12}+30t^{14}+107t^{16}+50t^{18} + \cdots + t^{36}}{(1-t^{8})^{3}(1-t^{4})^{3}(1-t^{2})^{6}}$ & $6t^2+9t^4+6t^6+8t^8-21t^{10}-31t^{12}+47t^{16}$\\
         \bottomrule
    \end{tabular}
    \caption{Unrefined Hilbert series for the Coulomb branches of several examples of the $BC$-chain quiver \eqref{quiv:appendix_BC_chain} labelled by $(N, k)$.}
    \label{table:BC_chain_HS}
\end{table}
\bibliographystyle{JHEP}
\bibliography{bibli.bib}

\providecommand{\href}[2]{#2}\begingroup\raggedright\begin{thebibliography}{10}

\bibitem{Cabrera:2018ann}
S.~Cabrera and A.~Hanany, \emph{{Quiver Subtractions}}, \href{http://dx.doi.org/10.1007/JHEP09(2018)008}{\emph{JHEP} {\bfseries 09} (2018) 008}, [\href{https://arxiv.org/abs/1803.11205}{{\ttfamily 1803.11205}}].

\bibitem{Gledhill:2021cbe}
K.~Gledhill and A.~Hanany, \emph{{Coulomb branch global symmetry and quiver addition}}, \href{http://dx.doi.org/10.1007/JHEP12(2021)127}{\emph{JHEP} {\bfseries 12} (2021) 127}, [\href{https://arxiv.org/abs/2109.07237}{{\ttfamily 2109.07237}}].

\bibitem{Bourget:2023dkj}
A.~Bourget, M.~Sperling and Z.~Zhong, \emph{{Decay and Fission of Magnetic Quivers}},  \href{https://arxiv.org/abs/2312.05304}{{\ttfamily 2312.05304}}.

\bibitem{Cabrera:2016vvv}
S.~Cabrera and A.~Hanany, \emph{{Branes and the Kraft-Procesi Transition}}, \href{http://dx.doi.org/10.1007/JHEP11(2016)175}{\emph{JHEP} {\bfseries 11} (2016) 175}, [\href{https://arxiv.org/abs/1609.07798}{{\ttfamily 1609.07798}}].

\bibitem{Cabrera:2017njm}
S.~Cabrera and A.~Hanany, \emph{{Branes and the Kraft-Procesi transition: classical case}}, \href{http://dx.doi.org/10.1007/JHEP04(2018)127}{\emph{JHEP} {\bfseries 04} (2018) 127}, [\href{https://arxiv.org/abs/1711.02378}{{\ttfamily 1711.02378}}].

\bibitem{Hwang:2021ulb}
C.~Hwang, S.~Pasquetti and M.~Sacchi, \emph{{Rethinking mirror symmetry as a local duality on fields}}, \href{http://dx.doi.org/10.1103/PhysRevD.106.105014}{\emph{Phys. Rev. D} {\bfseries 106} (2022) 105014}, [\href{https://arxiv.org/abs/2110.11362}{{\ttfamily 2110.11362}}].

\bibitem{Comi:2022aqo}
R.~Comi, C.~Hwang, F.~Marino, S.~Pasquetti and M.~Sacchi, \emph{{The SL(2, \ensuremath{\mathbb{Z}}) dualization algorithm at work}}, \href{http://dx.doi.org/10.1007/JHEP06(2023)119}{\emph{JHEP} {\bfseries 06} (2023) 119}, [\href{https://arxiv.org/abs/2212.10571}{{\ttfamily 2212.10571}}].

\bibitem{Giacomelli:2023zkk}
S.~Giacomelli, C.~Hwang, F.~Marino, S.~Pasquetti and M.~Sacchi, \emph{{Probing bad theories with the dualization algorithm. Part I}}, \href{http://dx.doi.org/10.1007/JHEP04(2024)008}{\emph{JHEP} {\bfseries 04} (2024) 008}, [\href{https://arxiv.org/abs/2309.05326}{{\ttfamily 2309.05326}}].

\bibitem{Giacomelli:2024laq}
S.~Giacomelli, C.~Hwang, F.~Marino, S.~Pasquetti and M.~Sacchi, \emph{{Probing bad theories with the dualization algorithm. Part II.}}, \href{http://dx.doi.org/10.1007/JHEP07(2024)165}{\emph{JHEP} {\bfseries 07} (2024) 165}, [\href{https://arxiv.org/abs/2401.14456}{{\ttfamily 2401.14456}}].

\bibitem{Grimminger:2024mks}
J.~F. Grimminger, W.~Harding and N.~Mekareeya, \emph{{Wreathing, discrete gauging, and non-invertible symmetries}}, \href{http://dx.doi.org/10.1007/JHEP01(2025)124}{\emph{JHEP} {\bfseries 01} (2025) 124}, [\href{https://arxiv.org/abs/2410.12906}{{\ttfamily 2410.12906}}].

\bibitem{Bennett:2024loi}
S.~Bennett, A.~Hanany, G.~Kumaran, C.~Li, D.~Liu and M.~Sperling, \emph{{Quiver Subtraction on the Higgs Branch}},  \href{https://arxiv.org/abs/2409.16356}{{\ttfamily 2409.16356}}.

\bibitem{Bennett:2024llh}
S.~Bennett, A.~Hanany and G.~Kumaran, \emph{{Orthosymplectic quotient quiver subtraction}}, \href{http://dx.doi.org/10.1007/JHEP12(2024)063}{\emph{JHEP} {\bfseries 12} (2024) 063}, [\href{https://arxiv.org/abs/2409.15419}{{\ttfamily 2409.15419}}].

\bibitem{beauville}
A.~Beauville, \emph{Symplectic singularities}, \href{http://dx.doi.org/10.1007/s002229900043}{\emph{Inventiones mathematicae} {\bfseries 139} (2000) 541--549}.

\bibitem{Bourget:2019aer}
A.~Bourget, S.~Cabrera, J.~F. Grimminger, A.~Hanany, M.~Sperling, A.~Zajac et~al., \emph{{The Higgs mechanism \textemdash{} Hasse diagrams for symplectic singularities}}, \href{http://dx.doi.org/10.1007/JHEP01(2020)157}{\emph{JHEP} {\bfseries 01} (2020) 157}, [\href{https://arxiv.org/abs/1908.04245}{{\ttfamily 1908.04245}}].

\bibitem{Bourget:2022ehw}
A.~Bourget, J.~F. Grimminger, A.~Hanany and Z.~Zhong, \emph{{The Hasse diagram of the moduli space of instantons}}, \href{http://dx.doi.org/10.1007/JHEP08(2022)283}{\emph{JHEP} {\bfseries 08} (2022) 283}, [\href{https://arxiv.org/abs/2202.01218}{{\ttfamily 2202.01218}}].

\bibitem{Bourget_2022dim6}
A.~Bourget and J.~F. Grimminger, \emph{Fibrations and hasse diagrams for 6d {SCFTs}}, \href{http://dx.doi.org/10.1007/jhep12(2022)159}{\emph{Journal of High Energy Physics} {\bfseries 2022} (dec, 2022) }.

\bibitem{Grimminger:2020dmg}
J.~F. Grimminger and A.~Hanany, \emph{{Hasse diagrams for 3d $ \mathcal{N} $ = 4 quiver gauge theories \textemdash{} Inversion and the full moduli space}}, \href{http://dx.doi.org/10.1007/JHEP09(2020)159}{\emph{JHEP} {\bfseries 09} (2020) 159}, [\href{https://arxiv.org/abs/2004.01675}{{\ttfamily 2004.01675}}].

\bibitem{Borokhov:2002ib}
V.~Borokhov, A.~Kapustin and X.-k. Wu, \emph{{Topological disorder operators in three-dimensional conformal field theory}}, \href{http://dx.doi.org/10.1088/1126-6708/2002/11/049}{\emph{JHEP} {\bfseries 11} (2002) 049}, [\href{https://arxiv.org/abs/hep-th/0206054}{{\ttfamily hep-th/0206054}}].

\bibitem{Borokhov:2002cg}
V.~Borokhov, A.~Kapustin and X.-k. Wu, \emph{{Monopole operators and mirror symmetry in three-dimensions}}, \href{http://dx.doi.org/10.1088/1126-6708/2002/12/044}{\emph{JHEP} {\bfseries 12} (2002) 044}, [\href{https://arxiv.org/abs/hep-th/0207074}{{\ttfamily hep-th/0207074}}].

\bibitem{Hanany:1996ie}
A.~Hanany and E.~Witten, \emph{{Type IIB superstrings, BPS monopoles, and three-dimensional gauge dynamics}}, \href{http://dx.doi.org/10.1016/S0550-3213(97)00157-0}{\emph{Nucl. Phys. B} {\bfseries 492} (1997) 152--190}, [\href{https://arxiv.org/abs/hep-th/9611230}{{\ttfamily hep-th/9611230}}].

\bibitem{Hanany:1999sj}
A.~Hanany and A.~Zaffaroni, \emph{{Issues on orientifolds: On the brane construction of gauge theories with SO(2n) global symmetry}}, \href{http://dx.doi.org/10.1088/1126-6708/1999/07/009}{\emph{JHEP} {\bfseries 07} (1999) 009}, [\href{https://arxiv.org/abs/hep-th/9903242}{{\ttfamily hep-th/9903242}}].

\bibitem{Hanany_2001}
A.~Hanany and J.~Troost, \emph{Orientifold planes, affine algebras and magnetic monopoles}, \href{http://dx.doi.org/10.1088/1126-6708/2001/08/021}{\emph{Journal of High Energy Physics} {\bfseries 2001} (Aug., 2001) 021–021}.

\bibitem{Feng_2000}
B.~Feng and A.~Hanany, \emph{Mirror symmetry by o3-planes}, \href{http://dx.doi.org/10.1088/1126-6708/2000/11/033}{\emph{Journal of High Energy Physics} {\bfseries 2000} (Nov., 2000) 033–033}.

\bibitem{2023arXiv231000521J}
D.~{Juteau}, P.~{Levy} and E.~{Sommers}, \emph{{Minimal special degenerations and duality}}, \href{http://dx.doi.org/10.48550/arXiv.2310.00521}{\emph{arXiv e-prints} (Sept., 2023) arXiv:2310.00521}, [\href{https://arxiv.org/abs/2310.00521}{{\ttfamily 2310.00521}}].

\bibitem{Generic_singularities}
B.~{Fu}, D.~{Juteau}, P.~{Levy} and E.~{Sommers}, \emph{{Generic singularities of nilpotent orbit closures}}, \href{http://dx.doi.org/10.48550/arXiv.1502.05770}{\emph{arXiv e-prints} (Feb., 2015) arXiv:1502.05770}, [\href{https://arxiv.org/abs/1502.05770}{{\ttfamily 1502.05770}}].

\bibitem{SOMMERS2001790}
E.~Sommers, \emph{Lusztig's canonical quotient and generalized duality}, \href{http://dx.doi.org/https://doi.org/10.1006/jabr.2001.8868}{\emph{Journal of Algebra} {\bfseries 243} (2001) 790--812}.

\bibitem{MARCUS198297}
N.~Marcus and A.~Sagnotti, \emph{Tree-level constraints on gauge groups for type i superstrings}, \href{http://dx.doi.org/https://doi.org/10.1016/0370-2693(82)90253-2}{\emph{Physics Letters B} {\bfseries 119} (1982) 97--99}.

\bibitem{Cabrera:2017ucb}
S.~Cabrera, A.~Hanany and Z.~Zhong, \emph{{Nilpotent orbits and the Coulomb branch of $T^\sigma (G)$ theories: special orthogonal vs orthogonal gauge group factors}}, \href{http://dx.doi.org/10.1007/JHEP11(2017)079}{\emph{JHEP} {\bfseries 11} (2017) 079}, [\href{https://arxiv.org/abs/1707.06941}{{\ttfamily 1707.06941}}].

\bibitem{Cabrera:2018ldc}
S.~Cabrera, A.~Hanany and R.~Kalveks, \emph{{Quiver Theories and Formulae for Slodowy Slices of Classical Algebras}}, \href{http://dx.doi.org/10.1016/j.nuclphysb.2018.12.022}{\emph{Nucl. Phys. B} {\bfseries 939} (2019) 308--357}, [\href{https://arxiv.org/abs/1807.02521}{{\ttfamily 1807.02521}}].

\bibitem{Bourget:2017tmt}
A.~Bourget and A.~Pini, \emph{{Non-Connected Gauge Groups and the Plethystic Program}}, \href{http://dx.doi.org/10.1007/JHEP10(2017)033}{\emph{JHEP} {\bfseries 10} (2017) 033}, [\href{https://arxiv.org/abs/1706.03781}{{\ttfamily 1706.03781}}].

\bibitem{Hanany:2016gbz}
A.~Hanany and R.~Kalveks, \emph{{Quiver Theories for Moduli Spaces of Classical Group Nilpotent Orbits}}, \href{http://dx.doi.org/10.1007/JHEP06(2016)130}{\emph{JHEP} {\bfseries 06} (2016) 130}, [\href{https://arxiv.org/abs/1601.04020}{{\ttfamily 1601.04020}}].

\bibitem{WITTEN1982324}
E.~Witten, \emph{{An $SU(2)$ Anomaly}}, \href{http://dx.doi.org/https://doi.org/10.1016/0370-2693(82)90728-6}{\emph{Physics Letters B} {\bfseries 117} (1982) 324--328}.

\bibitem{achar_lcq}
P.~N. Achar, \emph{An order-reversing duality map for conjugacy classes in lusztig's canonical quotient}, \href{http://dx.doi.org/10.1007/s00031-003-0422-x}{\emph{Transformation Groups} {\bfseries 8} (2003) 107--145}.

\bibitem{Bourget:2022tmw}
A.~Bourget and J.~F. Grimminger, \emph{{Fibrations and Hasse diagrams for 6d SCFTs}}, \href{http://dx.doi.org/10.1007/JHEP12(2022)159}{\emph{JHEP} {\bfseries 12} (2022) 159}, [\href{https://arxiv.org/abs/2209.15016}{{\ttfamily 2209.15016}}].

\bibitem{Hanany:2018vph}
A.~Hanany and G.~Zafrir, \emph{{Discrete Gauging in Six Dimensions}}, \href{http://dx.doi.org/10.1007/JHEP07(2018)168}{\emph{JHEP} {\bfseries 07} (2018) 168}, [\href{https://arxiv.org/abs/1804.08857}{{\ttfamily 1804.08857}}].

\bibitem{Cremonesi:2013lqa}
S.~Cremonesi, A.~Hanany and A.~Zaffaroni, \emph{{Monopole operators and Hilbert series of Coulomb branches of $3d$ $\mathcal{N} = 4$ gauge theories}}, \href{http://dx.doi.org/10.1007/JHEP01(2014)005}{\emph{JHEP} {\bfseries 01} (2014) 005}, [\href{https://arxiv.org/abs/1309.2657}{{\ttfamily 1309.2657}}].

\bibitem{Hanany:2016pfm}
A.~Hanany and M.~Sperling, \emph{{Algebraic properties of the monopole formula}}, \href{http://dx.doi.org/10.1007/JHEP02(2017)023}{\emph{JHEP} {\bfseries 02} (2017) 023}, [\href{https://arxiv.org/abs/1611.07030}{{\ttfamily 1611.07030}}].

\bibitem{juteau2023minimal}
D.~Juteau, P.~Levy and E.~Sommers, \emph{Minimal special degenerations and duality}, {\emph{arXiv preprint arXiv:2310.00521} (2023) }.

\bibitem{2023arXiv230807398F}
B.~{Fu}, D.~{Juteau}, P.~{Levy} and E.~{Sommers}, \emph{{Local geometry of special pieces of nilpotent orbits}}, \href{http://dx.doi.org/10.48550/arXiv.2308.07398}{\emph{arXiv e-prints} (Aug., 2023) arXiv:2308.07398}, [\href{https://arxiv.org/abs/2308.07398}{{\ttfamily 2308.07398}}].

\bibitem{Ferlito:2016grh}
G.~Ferlito and A.~Hanany, \emph{{A tale of two cones: the Higgs Branch of Sp(n) theories with 2n flavours}},  \href{https://arxiv.org/abs/1609.06724}{{\ttfamily 1609.06724}}.

\bibitem{Bourget:2023cgs}
A.~Bourget, J.~F. Grimminger, A.~Hanany, R.~Kalveks, M.~Sperling and Z.~Zhong, \emph{{A tale of N cones}}, \href{http://dx.doi.org/10.1007/JHEP09(2023)073}{\emph{JHEP} {\bfseries 09} (2023) 073}, [\href{https://arxiv.org/abs/2303.16939}{{\ttfamily 2303.16939}}].

\bibitem{Kapustin_1999}
A.~Kapustin and M.~J. Strassler, \emph{On mirror symmetry in three dimensional abelian gauge theories}, \href{http://dx.doi.org/10.1088/1126-6708/1999/04/021}{\emph{Journal of High Energy Physics} {\bfseries 1999} (Apr., 1999) 021–021}.

\bibitem{Zafrir:2015ftn}
G.~Zafrir, \emph{{Brane webs and $O5$-planes}}, \href{http://dx.doi.org/10.1007/JHEP03(2016)109}{\emph{JHEP} {\bfseries 03} (2016) 109}, [\href{https://arxiv.org/abs/1512.08114}{{\ttfamily 1512.08114}}].

\end{thebibliography}\endgroup
\end{document}